\documentclass[prd,twocolumn,reprint,preprintnumbers,nofootinbib,superscriptaddress]{revtex4-2}
\input{header.tex}

\begin{document}

\reportnum{-2}{CERN-TH-2025-073}\reportnum{-3}{FERMILAB-PUB-25-0200-CSAID}

\title{New physics particles mixing with mesons: production in the fragmentation chain}
\author{Yehor Kyselov}
\email{kiselev883@gmail.com}
\affiliation{Taras Shevchenko National University of Kyiv, Kyiv, Ukraine} 
\author{Stephen Mrenna}
\email{mrenna@fnal.gov}
\affiliation{Department of Physics, University of Cincinnati, Cincinnati, OH 45221, USA}
\affiliation{Fermi National Accelerator Laboratory, Batavia, IL 60510, USA}
\author{Maksym~Ovchynnikov}
\email{maksym.ovchynnikov@cern.ch}
\affiliation{Theoretical Physics Department, CERN, 1211 Geneva 23, Switzerland}

\date{\today}

\begin{abstract}

A class of extensions to the Standard Model adds hypothetical long-lived particles (LLPs) that have mass- or kinetic-mixing with neutral mesons, such as pions or rho mesons. The mixing can contribute significantly to the production of LLPs at proton accelerator experiments, and no consistent description of these production modes exists in the literature. In this paper, we develop a framework for studying different LLPs -- dark photons, vector mediators coupled to the baryon current, and axion-like particles with different coupling patterns. In particular, we implement the production mechanisms in \texttt{PYTHIA8}, study how the overall flux and kinematic distributions depend on the LLP's mass, and compare various sub-processes where the mixing contributes -- proton bremsstrahlung, meson decay, and production in the fragmentation chain.
We find that our new description of LLP production predicts an integrated flux that differs from current approaches by one to two orders of magnitude, and highlight the unavoidable theoretical uncertainties coming from poor knowledge of the properties of heavy mesons.

\end{abstract}

\maketitle

\section{Introduction}
\label{sec:introduction}

To account for several outstanding phenomena in particle physics and cosmology -- such as dark matter, neutrino oscillations, and the baryon asymmetry of the Universe -- the Standard Model (SM) likely requires an extension that includes hypothetical new particles. Their properties, including masses and interactions, are not strictly determined by current observations; therefore, viable scenarios span a mass range from below an eV to the Planck scale. From an experimental standpoint, it is particularly appealing to focus on masses $\lesssim \Lambda_{\text{EW}}$, notably near the GeV scale, where particles can be copiously produced and exhibit rich phenomenology. The viable parameter space typically entails small interaction strengths or, equivalently, long lifetimes -- hence the term Long-Lived Particles (LLPs)~\cite{Beacham:2019nyx,Antel:2023hkf,Bondarenko:2018ptm,Boiarska:2019jym,Ilten:2018crw,Kyselov:2024dmi}. Searches for LLPs form a central goal of ``intensity frontier'' experiments~\cite{Beacham:2019nyx,Antel:2023hkf}, including the currently running Belle II~\cite{Bertholet:2021hjl,Bernreuther:2022jlj}, the recently approved SHiP~\cite{SHiP:2015vad,Alekhin:2015byh,Aberle:2839677}, beam dump experiments at Fermilab~\cite{DUNE:2020lwj,Apyan:2022tsd}, new techniques at the main LHC detectors~\cite{Lee:2018pag,CMS:2024ake,Kholoimov:2025cqe}, and proposed facilities at the LHC, such as the Forward Physics Facility~\cite{Feng:2022inv} and off-axis detectors~\cite{Aielli:2019ivi,Bauer:2019vqk,MATHUSLA:2019qpy}.

``Minimal'' extensions of the SM that introduce an LLP $X$ involve gauge-invariant operators of lowest possible dimension, describing how $X$ interacts with SM fields~\cite{Alekhin:2015byh,Beacham:2019nyx,Antel:2023hkf}. Depending on $X$'s spin, LLPs can be scalar (often termed ``Higgs-like''~\cite{Boiarska:2019jym}), pseudoscalar (axion-like particles, or ALPs~\cite{Bauer:2020jbp}), fermionic (heavy neutral leptons, or HNLs~\cite{Bondarenko:2018ptm}), or vector particles (dark photons, $B-L$ mediators, etc.~\cite{Ilten:2018crw}). Qualitatively, Higgs-like scalars behave like light Higgs bosons; ALPs share features with $\pi^{0},\eta,\eta'$ mesons; HNLs resemble massive neutrinos; and vector LLPs may be likened to massive photons (dark photons) or vector mesons ($B-L$ mediators). These LLPs may couple to a hypothetical dark sector, such as dark matter~\cite{Beacham:2019nyx,Berlin:2018jbm,DallaValleGarcia:2023xhh}.

Bosonic LLPs often couple to quark or gluon bilinears. In the GeV mass range, these couplings can be re-expressed in terms of various hadronic bound states, such as neutral mesons $m^{0}$ or baryons. Among the resulting contributions in the effective Lagrangian, one finds terms involving both the LLP and a bound state, giving rise to a \emph{mixing} between the two. A simple example is the dark photon, which couples to the electromagnetic current; vector mesons like $\rho^{0},\omega,\phi$ appear as poles in this current~\cite{Sakurai:1960ju,Gell-Mann:1961jim,Kroll:1967it}.

Such mixing implies that the interaction eigenstate $m^{0}_{\text{int}}$ of the meson is a superposition of the meson's mass eigenstate $m^{0}$ and a small component of the LLP $X$, the latter being proportional to a \emph{mixing angle}, $\theta_{m^{0}X}$. Consequently, an analog of the process involving $m^{0}$ -- either as an external particle or a virtual state -- exists for $X$. As a result, often, the mixing with various mesons determines the LLP phenomenology at accelerator experiments, controlling production and decay modes. It is therefore essential to understand it.

Let us focus on the LLP production modes. Examples include meson decays (such as $\pi^{0}\to\gamma\rho^{*}\to\gamma X$), proton bremsstrahlung~\cite{Blumlein:2013cua}, where an incoming proton emits $X$ quasi-elastically, and hadronization processes in which a meson is replaced by $X$ at the final step of fragmentation. Previous studies have approximated the production rate of LLPs in some of these processes by multiplying the flux of the mother meson by $|\theta_{m^{0}X}|^{2}$~\cite{Berlin:2018jbm,Jerhot:2022chi}. Although useful as a rough estimate, this approach overlooks the mass dependence of the production flux and the detailed kinematics of LLPs. For instance, ALPs mix with $\pi^{0}$, which are copiously produced through both fragmentation and decays of other mesons (e.g., $\eta,\rho^{0},K$). While meson decays may efficiently produce light ALPs, they stop contributing once the ALP mass surpasses the decay threshold -- a feature that is not captured in the naive approximation. In addition, in the case of ALPs having coupling to gluons, the description may suffer from a deeper inconsistency -- it depends on unphysical chiral rotation parameters~\cite{Ovchynnikov:2025gpx}.

Despite these problems, the simplified procedure has gained traction in many tools for calculating LLP event rates~\cite{Kling:2021fwx,Jerhot:2022chi,Ovchynnikov:2023cry}, and the same shortcoming extends to various beyond-minimal scenarios where the LLPs serve as mediators to a hypothetical dark sector, such as dark matter~\cite{Berlin:2018jbm,Beacham:2019nyx,Antel:2023hkf,Fitzpatrick:2023xks,Foguel:2024lca}.

In this work, we refine the description of mixing-induced production channels by consistently treating both meson decays and hadronization. We implement these mechanisms in publicly available LLP event generators, namely \texttt{PYTHIA8}~\cite{Bierlich:2022pfr}\footnote{The implementation is available at \href{https://gitlab.com/YehorKyselyov/pythia-mixing/-/tree/dev}{https://gitlab.com/YehorKyselyov/pythia-mixing/-/tree/dev}.} and \texttt{SensCalc}~\cite{Ovchynnikov:2023cry},\footnote{Available on~\href{https://doi.org/10.5281/zenodo.7957784}{\texttt{Zenodo}}.} where they can be readily employed to determine experimental constraints and sensitivities.

The remainder of this paper is organized as follows. Sec.~\ref{sec:mixing-definition} defines meson mixing for LLPs and reviews the minimal models in which it arises. Sec.~\ref{sec:state-of-the-art} explores the production channels where the mixing contributes and surveys their approximate treatment in previous studies. Sec.~\ref{sec:approach} details our improved approach, while Sec.~\ref{sec:implementation} discusses implementation in \texttt{PYTHIA8} and \texttt{SensCalc}~\cite{Ovchynnikov:2023cry}. Sec.~\ref{sec:toy} investigates a toy model with a single mixing scenario and illustrates the qualitative impact on the LLP flux. Finally, Sec.~\ref{sec:case-studies} applies our framework to several scenarios, including vector mediators and ALPs, and we conclude in Sec.~\ref{sec:conclusions}.

\section{Mixing with mesons}
\label{sec:mixing-definition}

The generic pattern of the interactions leading to the mixing of an LLP $X$ with a meson $m^{0}$ is 
\begin{equation}
\mathcal{L} = X^{a}V_{a}, \quad \langle 0|V_{a}|m^{0}\rangle \neq 0,
\label{eq:lagrangian-mixing-pattern}
\end{equation}
where $a$ is the index corresponding to the number of spin degrees of freedom of $X$, while $V_{a}$ is the multi-field operator of quarks and gluons, with a non-zero matrix element between the one-meson state $|m^{0}\rangle$ and vacuum.

\begin{figure}[t!]
    \centering
    \includegraphics[width=0.9\linewidth]{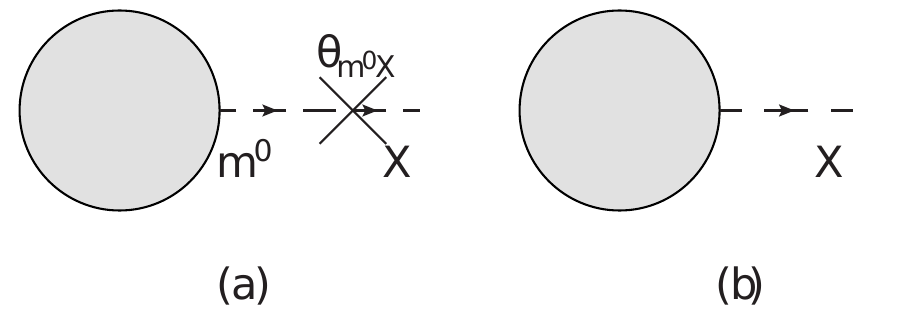}
    \caption{The structure of the matrix element of some arbitrary process involving an LLP $X$ mixing with an SM particle $m^{0}$. There are two contributions coming from the summands in the effective Lagrangian~\eqref{eq:operator-mixing}: from the terms describing mixing with $m^{0}$ (a), and from the multi-field operator $\mathcal{L}_{D\geq 3}$ (b). The first summand has the same structure as the matrix element with an external $m^{0}$ but with the replacement $m_{m^{0}}
    \to m_{X}$ and $m^{0}\to \theta_{m^{0}X} X$, where $\theta_{m^{0}a}$ is the mixing angle between $m_{0}$ and $X$ (Eq.~\eqref{eq:mixing-angle-definition}).}
    \label{fig:generic-matrix-element}
\end{figure}

The effective Lagrangian describing $X$'s interaction with mesons is thus
\begin{multline}
\mathcal{L} \subset \sum_{m^{0}}(M_{Xm^{0}}X^{a}m^{0a} + K^{\mu\nu}_{Xm^{0}}\partial_{\mu}X^{a}\partial_{\nu}m^{0a})\\ +\mathcal{L}_{D\geq 3},
\label{eq:operator-mixing}
\end{multline}
Here, the first two terms are quadratic in fields and describe the mixing, with $M_{Xm^{0}},K_{Xm^{0}}$ being constants; the first term represents the mass mixing, whereas the second one is the kinetic mixing. $\mathcal{L}_{D\geq 3}$ stands for the operators having the number of fields $\geq 3$ and including at least one $X$.

Let us now study how the mixing terms contribute to various interactions with $X$. Namely, consider some generic process $Y\to X+Y'$, where $Y,Y'$ are arbitrary SM states. There are two contributions to the matrix element, see Fig.~\ref{fig:generic-matrix-element}: 
\begin{equation}
\mathcal{M}_{Y \to  X+ Y'} = \sum_{m^{0}}\mathcal{M}^{\text{mixing}}_{m^{0}} + \mathcal{M}^{\text{direct}},
\label{eq:meson-decay-mixing-pattern}
\end{equation}
The first term (denoted by ``mixing'') comes from the quadratic operator~\eqref{eq:operator-mixing} with the given meson $m^{0}$. The second summand (denoted by ``direct'') originates from the non-quadratic terms.  

Assuming the mass mixing, the first summand takes the form
\begin{equation}
    \mathcal{M}^{\text{mixing}}_{m^{0}} =  \tilde{\mathcal{M}}^{i}\cdot D_{ij}(p_{X})M_{Xm^{0}}\epsilon^{j}(p_{X})
    \label{eq:matrix-element-mixing-contribution}
\end{equation}
Here, $\epsilon^{j}$ is the $X$'s polarization, $D_{ij}$ is the $m^{0}$'s propagator, while $\tilde{\mathcal{M}}$ stays for the rest of the matrix element. Independently of spin of $X$, we have $D_{ij}\epsilon^{j} \sim \epsilon_{i}/(m_{X}^{2}-m_{m^{0}}^{2}-i\Gamma_{m^{0}}m_{m^{0}})$, up to the $i$ factors. Therefore, the total matrix element becomes
$\mathcal{M} = \theta_{m^{0}X}\tilde{\mathcal{M}}^{i}\epsilon_{i}$, where we introduced \emph{the mixing angle}
\begin{equation}
    \theta_{m^{0}X}= \frac{M_{Xm^{0}}}{m_{X}^{2}-m_{m^{0}}^{2}-i\Gamma_{m^{0}}m_{m^{0}}}
    \label{eq:mixing-angle-definition}
\end{equation}
The mixing angle is enhanced resonantly around $m_{X} = m_{m^{0}}$ and amplifies $\mathcal{M}^{\text{mixing}}_{m^{0}}$, potentially making it the dominant term. This makes it crucial to understand thoroughly how the mixing contributes to the probabilities of various processes. Treating the kinetic mixing would be completely equivalent.

Further, instead of treating the quadratic Lagrangian as an interaction operator, we diagonalize it. This is more convenient because some models, such as ALPs, exhibit a non-trivial mixing structure with various mesons.

\subsection{Examples of LLPs with mixing}
\label{sec:model-examples}

Now, let us switch to realistic models where LLPs mix with mesons. As examples, we consider the lowest-dimensional, gauge-invariant interactions of scalar, pseudoscalar, and vector new physics particles, and discuss how mixing emerges. The mixing patterns of these LLPs are summarized in Table~\ref{tab:mixing}, while below, we provide details.

\begin{table*}[t!]
    \centering
    \begin{tabular}{|c|c|c|c|c|}
      \hline LLP & Effective Lagrangian & Mixing pattern \\ \hline
       ALP  & Eq.~\eqref{eq:lagrangian-alp} & \makecell{$\pi^{0},\eta,\eta'$ and their excitations}\\ \hline  
       Dark photon  & \makecell{Eq.~\eqref{eq:lagrangian-vectors} with $\alpha_{B} = 0$}  & \makecell{$\rho^{0},\omega,\phi$ and their excitations}\\ \hline
        $B-L_{\alpha}$ mediators  & \makecell{Eq.~\eqref{eq:lagrangian-vectors} with $\epsilon = 0$}  & \makecell{$\omega,\phi$  and their excitations} \\ \hline
         Higgs-like scalars    & Eq.~\eqref{eq:lagrangian-scalars} & $f_{0}(980)$ and their excitations\\ \hline
    \end{tabular}
    \caption{The list of the simple LLP models considered in this paper, the corresponding effective Lagrangians, and the pattern of the LLP mixing with various mesons. The results of the study may also be applied to various modifications of these simple models, such as those changing the mixing pattern, or adding different operators that do not affect the mixing~\cite{DallaValleGarcia:2025aeq,Bernreuther:2022jlj}.}
    \label{tab:mixing}
\end{table*}

\paragraph{Higgs-like scalars.} The Higgs portal model adds the following renormalizable interactions of a scalar particle $S$ with the Higgs doublet $H$~\cite{Beacham:2019nyx}:
\begin{equation}
    \mathcal{L} = \alpha_{1} SH^{\dagger}H + \alpha_{2}S^{2}H^{\dagger}H
    \label{eq:lagrangian-scalars}
\end{equation}
Here, $\alpha_{1,2}$ are interaction constants. Below the electroweak scale, the $SH^{\dagger}H$ term induces a mixing with the Higgs boson, leading to a Yukawa interaction between $S$ and quarks $q$~\cite{Boiarska:2019jym}: 
\begin{equation}
    \mathcal{L}_{\text{eff},S} \subset \theta S \sum_{f}y_{f}\bar{f}f, \quad y_{f} = \frac{m_{f}}{v_{h}},
\end{equation}
where $\theta$ is the mixing angle between the scalar and the Higgs boson. The quark Yukawa operators carry the same quantum numbers as scalar mesons, such as $f_{0}$ and its excitations~\cite{Kuroda:2019jzm}, which gives rise to their mass mixing with $S$. 

\paragraph{Pseudoscalar particles.} Consider a pseudoscalar $a$ (called Axion-Like Particle, or ALP) having the following interactions~\cite{Bauer:2017ris,Bauer:2020jbp}:
\begin{multline}
   \mathcal{L} =  c_{G} \frac{a}{f_{a}} \frac{\alpha_{S}}{4\pi}G^{a}_{\mu\nu}\tilde{G}^{a,\mu\nu} +\frac{\partial^\mu a}{f_{a}} \sum_Q \bar{\Psi}_Q \, \mathcal{C}_Q\, \gamma_\mu\, \Psi_Q \\ +\mathcal{L}_{\text{non-hadronic}}
   \label{eq:lagrangian-alp}
\end{multline}
Here, $G_{\mu\nu},\tilde{G}_{\mu\nu}$ are the gluon strength and its dual, while $\Psi_{Q}$ are quark field multiplets. $c_{G},f_{a},c_{Q}$ are interaction constants. Finally, $\mathcal{L}_{\text{non-hadronic}}$ denotes non-hadronic interactions of ALPs, such as with $SU(2)_{L},U(1)_{Y}$ gauge fields and leptons.

Depending on the coupling pattern, the second term in~\eqref{eq:lagrangian-alp} may include flavor-diagonal couplings
\begin{equation}
    \frac{\partial^\mu a}{f_{a}} \sum_Q \bar{\Psi}_Q \, \mathcal{C}_Q\, \gamma_\mu\, \Psi_Q \subset \frac{\partial_{\mu}a}{f_{a}}\sum_{q}c_{q}\bar{q}\gamma^{\mu}\gamma_{5}q+\text{other terms} 
    \label{eq:ALP-lagr}
\end{equation}
In principle, there may also be flavor-non-diagonal couplings of ALPs to quarks. They may be present already in the tree-level Lagrangian~\eqref{eq:lagrangian-alp}, and/or appear as a result of electroweak loop corrections generated by the diagonal terms~\eqref{eq:lagrangian-alp}~\cite{Bauer:2020jbp}. Additionally, they may genuinely arise as a result of ensuring renormalizability in the context of effective field theory~\cite{Chakraborty:2021wda,Bisht:2024hbs}. 

Such flavor-violating couplings may generate the mixing of ALPs with flavored mesons. For simplicity, we assume that there are no flavor-violating couplings, leaving this for future work.

To find the mixing pattern of ALPs with mesons, let us first perform the following chiral rotation of the quark fields:
\begin{equation}
q\to \mathcal{\xi}q, \quad \xi = \exp\left[-i\kappa_{q} \frac{a}{f_{a}}\gamma_{5}\right],
\label{eq:chiral-rotation}
\end{equation}
with $\kappa_{q}$ being an arbitrary diagonal matrix satisfying the condition $\text{Tr}[\kappa_{q}] = 1$. The rotation converts the $aGG$ term into a ($\kappa_{q}$-dependent) summand analogous to the second term in Eq.~\eqref{eq:lagrangian-alp}, and also modifies the quark mass term, $\bar{q}m_{q}q \to \bar{q}\xi m \xi q$. It is then straightforward to obtain the correspondence between the ALP Lagrangian and Chiral Perturbation Theory (ChPT)~\cite{Bauer:2020jbp}:
\begin{multline}
    \mathcal{L}_{\text{eff,a}} = \frac{f_{\pi}^{2}}{4}B_{0}\text{Tr}[\Sigma \hat{m}^{\dagger}_{q}+ \hat{m}_{q}\Sigma^{\dagger}]\\ +\frac{f_{\pi}^{2}}{2} \frac{\partial_{\mu}a}{f}\text{Tr}[(c_{q}+c_{G}\kappa_{q})(\Sigma D^{\mu}\Sigma^{\dagger}+\text{h.c.})]
    \label{eq:chpt-axion}
\end{multline}
Here, $f_{\pi}\approx 93\text{ MeV}$ is the pion decay constant, $B_{0}$ is the ChPT condensate, $\Sigma = \exp\left[\frac{2iP}{f_{\pi}}\right]$ is the matrix of the pseudoscalar meson nonet $P$,  $D_{\mu}$ is the covariant derivative, and
\begin{equation}
\hat{m}_{q} = \exp\left[-ic_{G}\kappa_{q}\frac{a}{f_{a}}\right]m_{q}\exp\left[-ic_{G}\kappa_{q}\frac{a}{f_{a}}\right],
\end{equation}
with $m_{q} = \text{diag}(m_{u},m_{d},m_{s})$.

Expanding Eq.~\eqref{eq:chpt-axion} in the lowest order of $P$ and $a$, one finds mixing terms between $a$ and the neutral mesons $P^{0}=\pi^{0},\eta,\eta'$ (see Appendix~\ref{app:mixing} for details). Because of the smallness of the $P$'s decay widths, in the domain $m_{a}\approx m_{P^{0}}$, the mixing angles are resonantly enhanced. It breaks perturbativity of the ALP-meson interaction~\cite{Bauer:2017ris,Bauer:2019vqk,DiLuzio:2024jip}. Similarly to the other studies, we restrict ourselves to considering the masses outside these domains. In the final figures~\ref{fig:production-ALP} showing the ALP production probabilities, we highlight the vicinities of $m_{P^{0}}$ where our analysis breaks down.

Adding the interactions with pseudoscalar excitations $\pi^{0}(1300),\eta(1295),\dots$ in the Lagrangian~\eqref{eq:chpt-axion} also allows describing their mixing with the ALPs, although there are sizeable theoretical uncertainties~\cite{Ovchynnikov:2025gpx}. 

Including both the contributions from direct operators and mixing to the matrix elements of various processes with ALPs (recall Fig.~\ref{fig:generic-matrix-element}) ensures that there is no dependence on components of $\kappa_{q}$-matrix from Eq.~\eqref{eq:chiral-rotation} -- only on the unambiguous quantity $\text{Tr}[\kappa_{q}] \equiv 1$~\cite{Bauer:2021wjo,Ovchynnikov:2025gpx}.

\begin{figure*}[t]
    \centering
    \includegraphics[width=0.9\textwidth]{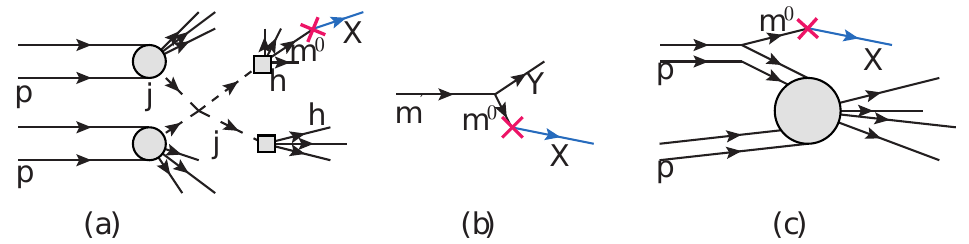}
    \caption{Diagrams of various production channels of a particle $X$ via the mixing with a neutral meson $m^{0}$: the fragmentation of partons in the deep-inelastic collisions (the diagram (a)), decays of heavy mesons $m^{'} \to X + Y$, with $Y$ being some Standard Model state (the diagram (b)), and proton bremsstrahlung (the diagram (c)). The red cross indicates the mixing vertex.}
    \label{fig:production-channels}
\end{figure*}

\paragraph{Vector mediators.} The simplest renormalizable and anomaly-free Lagrangians describing interactions of vector mediators includes the terms
\begin{equation}
    \mathcal{L} = V_{\mu}\big(e\epsilon J^{\mu}_{\text{EM}}+\sqrt{4\pi\alpha_{B}}(J^{\mu}_{\text{B}}-\sum_{\alpha}c_{\alpha}J^{\mu}_{L,\alpha})\big)
    \label{eq:lagrangian-vectors}
\end{equation}
Here, $J^{\mu}_{\text{EM}}, J^{\mu}_{\text{B}}$ are, respectively, electromagnetic and baryon currents:
\begin{align}
    J^{\mu}_{\text{EM}} &= \sum_{q}Q_{q}\bar{q}\gamma^{\mu}q+\sum_{l}Q_{l}\bar{l}\gamma^{\mu}l, \\  J^{\mu}_{\text{B}} &= \frac{1}{3}\sum_{q}\bar{q}\gamma^{\mu}q,
\end{align}
with $Q_{f}$ being the electric charge of the fermion $f$ in units of the proton electric charge. $J^{\mu}_{L,\alpha}$ is the lepton current of the flavor $\alpha = e,\mu,\tau$. The coefficients $c_{\alpha}$ are fixed such that there is no mixed SM anomaly~\cite{Dror:2017ehi}. 

The $V$ particle coupled solely to the EM current is called the dark photon (with the coupling $\epsilon \ll 1$). In the opposite case $\epsilon = 0$ and non-zero (yet tiny) $\alpha_{B}$, $V$ belongs to the family of the ``$B$ mediators'' coupled to the combination of the baryon and lepton currents~\cite{Alekhin:2015byh,Beacham:2019nyx,Ilten:2018crw}.

The EM current has a non-zero matrix element between the vacuum and 1-particle vector meson states, such as the ground states $\rho^{0},\omega,\phi$ as well as their excitations~\cite{OConnell:1995nse,Fujiwara:1984mp}; historically, this is known as vector meson dominance~\cite{Sakurai:1960ju,Kroll:1967it}. The baryon current, in contrast, only has non-zero matrix elements with $\omega,\phi$~\cite{Tulin:2014tya,Ilten:2018crw} and their excited states. As a result, the interaction~\eqref{eq:lagrangian-vectors} induces an effective mass mixing between $V$ and these mesons.

Explicit expressions for the mixing angles of ALPs, dark photons, and the mediators coupled to the baryon current may be found in Appendix~\ref{app:mixing}. For the ALPs, we also provide a comprehensive step-by-step derivation of the corresponding expressions, following Ref.~\cite{Ovchynnikov:2025gpx}. 

\subsection{LLP production channels via mixing}
\label{sec:production-modes}

There are several production processes of an LLP $X$ to which the mixing directly contributes (see Fig.~\ref{fig:production-channels}):
\begin{itemize}
    \item[--] During \textbf{\textit{parton fragmentation}} in deep-inelastic proton collisions: each time a parton fragments into $m^{0}$, there is a small probability (proportional to the mixing angle squared) that instead $X$ will appear.
      %The resulting $X$ typically occurs as final state radiation. 
    \item[--] Via \textbf{\textit{decays of mesons}} $m^{'}$: 
    \begin{equation}
     m^{'}\to X +\text{other}
    \end{equation}
    For example, consider an LLP $X$ mixing with $\pi^{0}$. Then, the existence of the decay $\eta \to \pi^{+}\pi^{-}\pi^{0}$ automatically implies $\eta \to \pi^{+}\pi^{-}X$. Another illustration is the decay process $\pi^{0}\to 2\gamma$. Because of the mixing of $\rho^{0}$ with $\gamma$, it can be equivalently described by $\pi^{0} \to 2\rho^{0*} \to 2\gamma$~\cite{Fujiwara:1984mp}. Therefore, any vector $X$ that has a mixing with $\rho^{0}$ can be produced in the process $\pi^{0}\to \gamma X$.
    \item[--] In \textbf{\textit{proton bremsstrahlung}}, which is an approximation of the production of $X$ as initial state radiation where the $X$ particle gets emitted as the initial state radiation from the elastic $ppX$ vertex~\cite{Boiarska:2019jym}, in a sub-process
    \begin{equation}
        p\to p'+V,
        \label{eq:quasi-real-emission}
    \end{equation}
    with $p'$ being a quasi-real proton with non-zero virtuality $p_{p'}^{2}-m_{p}^{2}$. The mixing contributes via inserting the internal $m^{0}$ lines in the effective $ppX$ vertex, and results in the resonant enhancement of the bremsstrahlung amplitude via the elastic $ppX$ form-factor around the masses of $m^{0}$. Using the so-called unitary and analytic model of extended vector meson dominance~\cite{Adamuscin:2016rer,Foroughi-Abari:2024xlj}, the latter may be represented in a form
    \begin{equation}
      F_{ppX}(q) = \sum_{m^{0}}f_{m^{0}}\frac{m_{m^{0}}^{2}}{q^{2}-m_{m^{0}}^{2}-i\Gamma_{m^{0}}m_{m^{0}}},
      \label{eq:form-factor-intro}
    \end{equation}
    where $q$ is the momentum transfer, the $f_{m^{0}}$ are phenomenological expansion constants, and the $\Gamma_{m^{0}}$ are decay widths.
\end{itemize}
Often, these modes are the main production mechanisms of new physics particles with mass in the GeV range~\cite{Beacham:2019nyx,Antel:2023hkf,Jerhot:2022chi,Ovchynnikov:2023cry}. However, depending on the LLP, they are described at different levels of sophistication in the literature. We summarize the situation below.

\section{Description of production channels: state-of-the-art}
\label{sec:state-of-the-art}

\subsection{Vector particles}
\label{sec:state-of-the-art-vector}
Consider first the case of vector $X$s -- dark photons and the mediators coupled to the baryon current. The contributions from decays, hadronization, and bremsstrahlung are considered exclusively~\cite{Beacham:2019nyx}. 

The production by decays of mesons is represented by the processes $P^{0}\to X\gamma$, where $P^{0}  =\pi^{0},\eta,\eta'$ is a pseudoscalar meson (recall Table~\ref{tab:decay-processes}), and $\omega \to \pi^{0}V$. The branching ratios of these decays can be obtained using the Hidden Local Symmetry approach of vector meson dominance~\cite{Ilten:2018crw}. It is well-understood, with the theoretical uncertainty on the production mode being within 20\%. There is an additional uncertainty coming from the uncertainty of our description of the flux of mesons; we discuss it in Sec.~\ref{sec:uncertainty-our-approach}.

The production in fragmentation is often approximated by the relation~\cite{Berlin:2018jbm,Kling:2021fwx,Ovchynnikov:2023cry}
\begin{equation}
    \frac{d^{2}N_{V}^{\text{fragm}}}{d\theta_{V}\,dE_{V}}
    \;=\;
    \sum_{m^{0}}
    \bigl|\theta_{m^{0}V}\bigr|^{2}\,
    \frac{d^{2}N_{m^{0}}}{d\theta_{m^{0}}\,dE_{m^{0}}},
    \label{eq:DP-hadronization-approximation}
\end{equation}
where $m^{0}=\rho^{0},\omega,\phi$ are the vector mesons mixing with $X$: \textit{i.e.}, the flux of $V$\!s is given by the flux of the mother mesons times the squared mixing angle.

Beyond the overall factor $\theta_{m^{0}V}$, this approximation neglects the influence of the $V$ mass on both the kinematics and the total flux. Intuitively, for $m_{V}<m_{m^{0}}$, one expects a broader angular distribution and a softer energy spectrum, as it is relatively straightforward to produce lighter $V$ as a secondary particle in the fragmentation process. Conversely, heavier $V$ states should yield a narrower angular spread and a harder energy distribution, leading to a suppression of their total production. Such mass-dependent effects become particularly important at facilities operating with lower center-of-mass collision energies. 

Finally, consider proton bremsstrahlung. For dark photons, the state-of-the-art description from~\cite{Blumlein:2013cua} extrapolates a quasi-elastic treatment of the bremsstrahlung process to the full inelastic scattering regime. However, it does not account for the theoretical uncertainties stemming from the limited virtuality of the intermediate proton after emitting the dark photon (see diagram (c) in Fig.~\ref{fig:production-channels}) or from the variations in the resonance parameters that define the effective form factor in the vertex $Xpp$, Eq.~\eqref{eq:form-factor-intro}. The latter can be determined using measurements of the rates of the electromagnetic processes in the two ``physical'' regions of momentum transfer, namely $t = (p_p - p_{p'})^2 < 0$ (using $ep\to ep$ scattering) and $t > 4m_p^2$ (using the $e^{+}e^{-}\leftrightarrow p\bar{p}$ process). For proton bremsstrahlung producing an $X$ particle, $t = m_X^2>0$, so it is necessary to know the behavior of the form factor over the continuous interval $t > 0$. Consequently, an extrapolation into the region $0 < t < 4m_p^2$ is required. Any extrapolation must satisfy the conditions of analyticity and unitarity.

These issues have been recently explored in Ref.~\cite{Foroughi-Abari:2024xlj}, which combined the ``quasi-real'' description of the bremsstrahlung from Refs.~\cite{Altarelli:1977zs,Boiarska:2019jym} with the adjustments required to avoid unphysical longitudinal divergences. To calculate the form-factor~\eqref{eq:form-factor-intro}, Ref.~\cite{Foroughi-Abari:2024xlj} employed the extended vector meson dominance model from~\cite{Adamuscin:2016rer}. This model uses a pole expansion in terms of vector mesons and their excitations, performs an analytic continuation to positive $t$, and determines the expansion coefficients using QCD sum rules and experimental data.

These combined uncertainties, from the variation of the proton virtuality in the quasi-real emission process $p\to p'+V$, Eq.~\eqref{eq:quasi-real-emission}, and the form factor, significantly affect the excluded dark photon parameter space and the sensitivities of future experiments~\cite{Kyselov:2024dmi} (we will return to this issue in Sec.~\ref{sec:challenges}). 

The situation is even more complicated for the $B-L_{\alpha}$ mediators. No data in the ``physical'' region is available for the form-factor, and the existing production descriptions (see, e.g.,~\cite{Boyarsky:2021moj}) violate analyticity and unitarity. 

\subsection{Axion-like particles}
\label{sec:state-of-the-art-alp}
For the pseudoscalar case (ALPs), the widely adopted description of the production modes~\cite{Jerhot:2022chi,Antel:2023hkf} is to approximate the contributions from all the channels from Fig.~\ref{fig:production-channels} by the expression
\begin{equation}
    \frac{d^{2}N_{X}^{\text{total}}}{d\theta_{X}dE_{X}} = \sum_{m^{0}}|\theta_{m^{0}X}|^{2}\frac{d^{2}N_{m^{0}}}{d\theta_{m^{0}}dE_{m^{0}}}\bigg|_{\theta_{m^{0}},E_{m^{0}}\to \theta_{X},E_{X}}
    \label{eq:state-of-the-art-alps}
\end{equation}
where $\theta_{m^{0}},E_{m^{0}}\to \theta_{X},E_{X}$ is some transformation relating the meson kinematics to the ALP
kinematics: \textit{i.e.}, the flux of LLPs is given by the fluxes of the corresponding pseudoscalar mesons times the squared mixing angles.

Using approximation~\eqref{eq:state-of-the-art-alps} for all the production modes introduces a few additional problems. The first challenge arises because the three production mechanisms shown in Fig.~\ref{fig:production-channels} have distinctly different dependencies on mass. Indeed, for example, the dominant contribution to the flux of SM pions $\pi^{0}$ (and hence ALPs with mass $m_{a}\simeq m_{\pi^{0}}$) comes from decays of heavier states, see Appendix~\ref{app:pion-production-from-decays}. However, as the mass of $X$ increases toward the kinematic threshold of these decays, the production probability drops sharply. In addition, the scaling of the decay branching ratios with $m_{a}$ can differ significantly from the naive expectation obtained by replacing the mass of the corresponding meson (e.g., $\pi^{0}$, $\eta$, $\eta'$) with the ALP mass in the decay matrix element. A notable example is the decay $K \to a + \pi$, where the dominant contribution comes from an operator that is sub-dominant in the corresponding SM decay $K\to 2\pi$~\cite{Gori:2020xvq,Bauer:2021mvw}.

The second problem is related to the transformation $\theta_{m^{0}},E_{m^{0}}\to \theta_{X},E_{X}$. It is obtained by going to the collision center-of-mass frame, replacing the meson mass with the ALP mass, and boosting back. The transformation is ambiguous -- it requires conserving either energy \textit{or} momentum in the CM frame (see Appendix of Ref.~\cite{Jerhot:2022chi}). In addition, it introduces unphysical bumps in the angular distribution in the case of a small invariant mass of the collisions (e.g., for the energies where the European Spallation Source~\cite{Alekou:2022emd} is sensitive to ALPs).

Finally, the third problem is the dependence of the description~\eqref{eq:state-of-the-art-alps} on the unphysical chiral rotation from Eq.~\eqref{eq:chiral-rotation}, namely on the components of the chiral rotation matrix $\kappa_{q}$ other than the unambiguous combination $\text{Tr}[\kappa_{q}] = 1$. The issue is encapsulated in the mixing angles $\theta_{m^{0}X}$, which are explicitly $\kappa_{q}$-dependent. To maintain consistency of the approach, it is necessary to add the direct contribution from multi-field operators resulting from the ALP Lagrangian~\eqref{eq:lagrangian-alp} to the production processes.

Some progress has been made on these issues in the literature. Reference~\cite{Blinov:2021say} treated proton bremsstrahlung and certain meson decays in a manner that ensured independence on $\kappa_{q}$, but did not consider production through fragmentation, account for theoretical uncertainties in bremsstrahlung, or include the contributions of various intermediate states in meson-decay matrix elements. Reference~\cite{Aielli:2019ivi} did include production via fragmentation using a method similar to that in this work (see Sec.\ref{sec:approach}), but it did not eliminate the unphysical dependence, nor did it provide a detailed discussion of the implementation.

Finally, the recent study~\cite{Ovchynnikov:2025gpx}, which systematically revised ALP production modes, relies on the results presented in this paper.

\subsection{Higgs-like scalars}

At most experiments, the dominant production modes of Higgs-like scalars are decays of $B$ mesons and Higgs bosons~\cite{Boiarska:2019jym}; the direct production mechanisms such as proton bremsstrahlung are mainly important at facilities with low collision energy, such as Fermilab, where the yield of $B$ mesons is significantly suppressed. Apart from this, currently, there is a controversy regarding the nature of scalar mesons such as $f_{0}(980)$ -- whether it is a 2-quark or a 4-quark bound state, or a $K\bar{K}$ molecular structure~\cite{Chen:2003za,Achasov:2020aun,Ahmed:2020kmp}. This introduces ambiguity in the description of the mixing of scalars with the family of scalar mesons. As a result, it is currently impossible to calculate reliably the production yield from bremsstrahlung and fragmentation.

\section{Our approach}
\label{sec:approach}

In this section, we describe our method to calculate the LLP flux by various production modes. We start with discussing theoretical challenges in the description, which come from limited knowledge of the spectroscopy of mesons with mass $M\gtrsim 1\text{ GeV}$. Then, we proceed to calculate the matrix elements of decays of mesons for various LLPs, define the rate of production in fragmentation, and obtain the flux of LLPs produced by proton bremsstrahlung.

\subsection{Theoretical challenges}
\label{sec:challenges}

As we mentioned in Sec.~\ref{sec:mixing-definition}, LLPs in realistic models mix not with individual mesons but rather with ``families'' of mesons sharing the same quantum numbers. These families include the ground state $m^{0}_{\text{ground}}$ (e.g., $\pi^{0}$) and its excitations $m^{0}_{\text{excitation}}$ (such as $\pi^{0}(1300)$, etc.). The ground states are well-understood both in terms of interactions and their parameters -- mass and partial decay widths. However, the excitations are not well understood~\cite{ParticleDataGroup:2024cfk}. Not only are their external properties poorly known, but sometimes even their internal nature is not clear; e.g., whether some of the pseudoscalar excitations are a 2-quark or a 4-quark bound state remains ambiguous~\cite{Giacosa:2024epf}. 

This imposes an unavoidable source of theoretical uncertainty in describing the production of LLPs. We summarize the impact of the uncertainty below: 
\begin{itemize}
    \item[--] \textit{Proton bremsstrahlung}. In Sec.~\ref{sec:production-modes}, we mentioned a crucial ingredient in calculating the bremsstrahlung flux -- knowing the elastic $ppX$ form-factor mediating the emission $p\to X+p'$~\eqref{eq:form-factor-intro}. It may be possible to fix the values of $f_{m^{0}}$ (see, e.g.,~\cite{Faessler:2009tn,Adamuscin:2016rer}) by an analytic extrapolation into the domain of positive $q^{2} = m_{X}^{2}$ supplemented with QCD sum rules and experimental data. 
    
    The uncertainties arise because of measurement errors in the EM scattering and, more importantly, the masses and widths of mesons $m^{0}$. Varying the latter, at the given point $q^{2} = m_{X}^{2}<4m_{p}^{2}$ in the ``unphysical'' region, there can be cancellations between the individual summands of~\eqref{eq:form-factor-intro} without spoiling the fit to the experimental data. This effect can increase the uncertainties by up to a few orders of magnitude.

    It is also challenging to minimize the uncertainty on the virtualities of the intermediate ``quasi-real'' proton $p'$ in the sub-process $p\to p'+V$ (recall the discussion around Eq.~\eqref{eq:quasi-real-emission}). The combined prediction of the LLP flux by bremsstrahlung and production in fragmentation can be compared to the experimental data on the inclusive production of vector mesons (in the case when the LLP resembles a vector meson); however, it is not clear how to disentangle the two contributions. In addition, this comparison would not tell anything about the proton form-factor.
    \item[--] \textit{Production in fragmentation}. The production flux of heavy meson excitations such as $\pi^{0}(1300)$, etc. (which are mainly produced in the fragmentation chain) has not even been measured. This is due to their large decay width and the subsequent complexity of extracting the information on their decay modes in a dense background environment. Because of this, it is not possible to include reliably the contribution of the mixing with these excitations to the produced flux of LLPs. 
    \item[--] \textit{Decays of mesons}. For the LLP models we consider, decays of mesons are almost free from uncertainty. This is because the branching ratios of decays are typically sizeable only for the lightest mesons. As a result, the main uncertainty comes from the unrelated source -- the error in the calculation of the fluxes of these mesons.
\end{itemize}

The same issue exists in the case of decays of LLPs.  The poor status of the exploration of the sector of pseudoscalar excitations results in a highly ambiguous description of their interaction with ALPs. For instance, this prevents us from calculating the hadronic decay width of ALPs and Higgs-like scalars in the mass range $m_{a}\simeq 1\text{ GeV}$. The latter is important for matching the two descriptions of the ALP decay width -- using perturbative QCD and the exclusive approach (i.e., including all the interactions with mesons)~\cite{Blackstone:2024ouf,Ovchynnikov:2025gpx}. 

To manage these complexities, we proceed in the following way. For the production in fragmentation, we only include the lowest excitations, restricted by the masses of mesons $M<1\text{ GeV}$. This way, our estimates are conservative. For the production by proton bremsstrahlung, we consider the state-of-the-art descriptions of the form-factors for various LLPs from Refs.~\cite{Blinov:2021say,Foroughi-Abari:2024xlj}. We comment on their limitations in Sec.~\ref{sec:bremsstrahlung}. Further, we highlight the status of the LLP phenomenology in our final plots, Figs.~\ref{fig:production-dark-photon}-\ref{fig:production-ALP}.

\subsection{Meson decays}
\label{sec:meson-decays}

We start by determining which mixings of $X$ would contribute to various decay processes of vector and pseudoscalar mesons, which are produced the most abundantly at proton accelerator experiments. For this purpose, we use the Chiral Perturbation Theory~\cite{Weinberg:1996kr} and extend it by adding vector mesons with the help of the Hidden Local Symmetry description of Vector Meson Dominance~\cite{Fujiwara:1984mp}. 

\begin{table*}[]
    \centering
    \begin{tabular}{|c|c|c|c|c|c|c|}
      \hline Process & SM counterpart & Contributed mixings \\ \hline
       $\rho^{\pm}\to \pi^{\pm}X$  & $\rho^{\pm}\to \pi^{\pm}\pi^{0}$ & $\pi^{0}$ \\ \hline $\eta\to \pi^{+}\pi^{-}X$& $\eta\to \pi^{+}\pi^{-}\pi^{0}$ & $\pi^{0},\eta,\eta'$ \\ \hline $\eta\to 2\pi^{0}X$& $\eta\to 3\pi^{0}$ & $\pi^{0},\eta,\eta'$ \\ \hline $K\to \pi X$ & $K\to \pi\pi$ & $\pi^{0},\eta,\eta'$ \\ \hline  $\omega\to X\gamma$ & $\omega\to \pi^{0}\gamma$ & $\pi^{0},\eta,\eta',\rho^{0}$ \\ \hline $\omega \to \pi^{+}\pi^{-}X$  & $\omega\to \pi^{+}\pi^{-}\pi^{0}$ & $\pi^{0},\eta,\eta'$ \\ \hline $\pi^{0}\to \gamma X$ & $\pi^{0}\to 2\gamma$ & $\omega,\rho^{0}$ \\ \hline $\eta\to \gamma X$ & $\eta\to 2\gamma$ & $\omega,\rho^{0}$ \\ \hline $\eta'\to \gamma X$ & $\eta'\to 2\gamma$ & $\omega,\rho^{0}$ \\ \hline $\omega\to \pi^{0}X$ & $\omega\to \pi^{0}\gamma$ & $\rho^{0},\omega$ \\ \hline
    \end{tabular}
    \caption{List of the most important decay processes of light mesons that can produce the $X$ particles via their mixings with neutral mesons. The columns are as follows: the $X$ production process, its Standard model counterpart, and the list of mixings that contribute. The table has been obtained using the ChPT Lagrangian supplemented by the interactions with vector mesons obtained within the framework of hidden local symmetry~\cite{Fujiwara:1984mp}.}
    \label{tab:decay-processes}
\end{table*}

For the relevant decay channels, we utilize the most frequent SM decay modes. The results are summarized in Table~\ref{tab:decay-processes}. The other decay channels are either too rare or have too limited phase space (an example is the decay $\Lambda^{+}\to pX$ with the SM analog $\Lambda^{+}\to p\pi^{0}$).

For calculating the branching ratios of various processes, we have to specify the LLP model: not only because it provides the mixing pattern with different mesons, but also, as discussed around Fig.~\ref{fig:generic-matrix-element} and in~\ref{sec:state-of-the-art-alp}, the contributions may involve not only the mixing part but also the direct contribution unrelated to the mixing.

First, consider the case of the vector mediators (denoted $V$). The $V$'s interaction operators, obtained after the insertion of the mixing, look exactly like as the operators of the interaction of the corresponding vector mesons.  The main decays in this case are of pseudoscalar mesons $P^{0} (\pi^{0},\eta,\eta'$) into a photon and $V$, occurring through a mixing of the photon with $\rho^{0}$, $\omega$, $\phi$ mesons.

For dark photons, the branching ratios can be expressed in terms of the branching ratio of the SM counterpart $P^{0}\to 2\gamma$ as~\cite{Ilten:2018crw}
\begin{equation}
    \text{Br}^{\text{DP}}_{P^{0}\to V\gamma} = 2\epsilon^{2} \text{Br}_{P^{0}\to 2\gamma}\times f(m_{V})\,.
    \label{eq:br-DP}
\end{equation}
The factor of 2 arises because there are not two identical particles in the final state. The phase space function $f(m_{V})$ follows from the calculation using the effective pion decay Lagrangian and reads
\begin{equation}
    f(m_{V}) \approx \left(1-\frac{m_{V}^{2}}{m_{P^{0}}^{2}}\right)^{3}\,.
\end{equation}
The branching ratios of the mediators coupled to the baryon current can be obtained by relating to Eq.~\eqref{eq:br-DP} using the $SU(3)$ representation of $V$ and mesons~\cite{Tulin:2014tya}:
\begin{equation}
    \text{Br}^{B-L_{\alpha}}_{P^{0}\to V\gamma} = \frac{\alpha_{B}}{\epsilon^{2}}\cdot \left(\frac{1}{3}\frac{\text{Tr}[T_{P^{0}}Q]}{\text{Tr}[T_{P^{0}}Q^{2}]}\right)^{2}\cdot \text{Br}^{\text{DP}}_{P^{0}\to V\gamma},
    \label{eq:br-BL}
\end{equation}
where $T_{P^{0}}$ is the generator of the decaying meson, and $Q = \text{diag}\left(\frac{2}{3},-\frac{1}{3}, -\frac{1}{3}\right)$ is the quark electric charge generator.

For the ALP case, we adopt the approach of Ref.~\cite{Ovchynnikov:2025gpx}. Namely, considering Eq.~\eqref{eq:chpt-axion}, we work in the three-flavor scenario (such that $\eta,\eta'$ are included), utilize the interaction Lagrangian including pseudoscalar, scalar, and vector mesons (relevant for calculating decay widths of mesons $\pi^{0},\eta,\eta'$ into ALPs), and add the flavor-violating operator inducing the $s\to d$ transition from~\cite{Bauer:2021wjo}.\footnote{Additional interactions induced by the Wess-Zumino-Witten terms may also contribute to the decays of vector mesons into ALPs, see~\cite{Bai:2024lpq}.} Furthermore, we also consider the $\mathcal{O}(\delta)$ limit, where $\delta = (m_{d}-m_{u})/(m_{d}+m_{u})$ is the isospin breaking parameter.

\subsection{Hadronization}
\label{sec:production-fragmentation}

Due to their mixing with mesons, LLPs may be thought of as ``exotic hadrons'' and hence can be produced in the hadronization process itself. In the Lund string model~\cite{Andersson:1983ia,Sjostrand:1984ic}, a
color singlet state of QCD partons fragments iteratively into hadrons.  At each step, the string end $q$ is combined with a parton $\bar Q$ to form a hadron $h$ ($q\bar Q$) with transverse momentum and a light-cone momentum fraction $z$. The fraction $z$ is selected based on the fragmentation function:
\begin{equation}
    f_{h}(z) = \frac{(1-z)^{a}}{z}\exp\left[-\frac{b}{z}(m_{h}^{2}+p_{T}^{2})\right]\,.
    \label{eq:fragmentation-function}
\end{equation}
In this expression, $m_h$ is the mass of the hadron produced by combining the string end with the new parton $\bar Q$, $p_T$ is the scalar sum of the string end and new hadron transverse momentum (selected from a Gaussian with a width fit to data), and $a$ and $b$ are parameters of the model that are fit to data.
The overlap of the $q\bar Q$ state with the LLP $X$ (aka exotic meson) is treated in the same way as for the $\pi, \rho, \eta$, \textit{etc.}

In this approach, since the exact exotic hadron mass is used, there is no ambiguity in the selection of the exotic hadron kinematics, and the overall energy and momentum of the event are conserved.

Similarly to the case of LLP production through meson decays, we must specify the LLP model. On top of that, to maintain the consistency of the description, we need to define the ``generalized'' mixing angle governing production in the fragmentation, $\Theta_{m^{0}X} = \theta_{m^{0}X}+\dots$, including the contribution from the mixing and the direct interaction operators. 

For the vector mediators, the interaction pattern with mesons is simple and can be described by the mixing only. This is not the case for the ALPs, given the complicated structure of the ChPT governing its interactions with pseudoscalar mesons. The latter is essential to maintain the consistency of the approach, as discussed in Sec.~\ref{sec:state-of-the-art-alp}.

We define $\Theta_{m^{0}a}$ for the ALPs in the following way. First, we consider some ``simple'' process involving the ALP and the mixing with the given meson $m^{0}$ and calculate its amplitude using the consistent Lagrangian with ALPs and mesons from Ref.~\cite{Ovchynnikov:2025gpx}. It can be expressed in the form
\begin{equation}
    \mathcal{M}_{X\to Y} = \Theta_{m^{0}a}\cdot \tilde{M}_{X\to Y},
\end{equation}
where 
\begin{equation}
\Theta_{m^{0}a} = \theta_{m^{0}a}+\sum_{m^{0'}\neq m^{0}}c_{m^{0'}}\theta_{m^{0'}a}+\Delta(\kappa)
\label{eq:theta-eff}
\end{equation}
is $\kappa$-independent effective coupling, with the first two terms coming from the mixing, and the last term from the direct interaction operator. Second, we extrapolate this coupling on the generic fragmentation chain. Details on the calculation of $\Theta_{m^{0}a}$ are given in Appendix~\ref{app:alp-fragmentation}.

\subsection{Proton bremsstrahlung}
\label{sec:bremsstrahlung}

We follow Refs.~\cite{Foroughi-Abari:2024xlj,Kyselov:2024dmi} for the description of the proton bremsstrahlung for dark photons. As in these works, we introduce the proton virtuality by the multiplicative form factor
\begin{equation}
\mathcal{F}_{V} = \frac{1}{1+\left(\frac{p'^{2}-m_{p}^{2}}{\Lambda_{p}^{2}}\right)^{2}},
\label{eq:reality-condition}
\end{equation}
where $p'$ is the momentum of the intermediate ``quasi-real'' proton. We vary the $\Lambda_{p}$ scale, defining the range of the virtuality of the intermediate proton $p'$ in the sub-process $p\to p' + V$, within the range $0.5\text{ GeV}<\Lambda_{p}<2\text{ GeV}$ (see a discussion in~\cite{Kyselov:2024dmi}).

The elastic proton form-factors have been calculated in~\cite{Foroughi-Abari:2024xlj} taking into account uncertainties in measurements of masses of meson excitations. The variation of the widths is ignored, motivated by the fact that it blows up the uncertainties.

We leave the discussion on the proton bremsstrahlung in the case of the mediators coupled to the baryon current for future work~\cite{Kyselov:2025ta}.

As for the ALPs, we adopt the description of the bremsstrahlung from~\cite{Blinov:2021say,Ovchynnikov:2025gpx}, using the same variation of the proton virtuality as in the dark photon case. The proton form-factor used in~\cite{Blinov:2021say} does not account for the mixing with pseudoscalar excitations, which may potentially underestimate its magnitude for the ALP mass range $m\gtrsim 1 \text{ GeV}$. 

\section{Implementation in \texttt{PYTHIA8} and \texttt{SensCalc}}
\label{sec:implementation}
In this section, we implement the approach described above in two tools: \texttt{PYTHIA8}\footnote{Available at \url{https://gitlab.com/YehorKyselyov/pythia-mixing/-/tree/dev}.} and \texttt{SensCalc}~\cite{Ovchynnikov:2023cry}\footnote{Available on~\href{https://doi.org/10.5281/zenodo.7957784}{\texttt{Zenodo}}.}. The former can be used to simulate the production of LLPs at various facilities, whereas the latter utilizes the production fluxes to calculate the event rate with LLPs at different experiments, taking into account their geometry and selection criteria on decay products.

We consider the following four models:
\begin{enumerate}
    \item Dark photons~\cite{Ilten:2018crw}.
    \item Mediators coupled to the baryon current~\cite{Ilten:2018crw}.
    \item ALPs universally coupled to fermions ($c_{q} = 1, c_{G} = 0$ in Eq.~\eqref{eq:lagrangian-alp}) at the scale $\Lambda = 1\text{ TeV}$~\cite{DallaValleGarcia:2023xhh}.
     \item ALPs coupled solely to gluons ($c_{q} = 0, c_{G} = 1$) at the scale $\Lambda = 1\text{ TeV}$~\cite{Aloni:2018vki,Blinov:2021say,Ovchynnikov:2025gpx}.
\end{enumerate}
For the ALPs, specifying the scale $\Lambda$ is important because the renormalization group flow from $\Lambda$ down to scales of interest $Q \simeq m_{\text{ALP}}$ modifies the coupling pattern~\cite{Bauer:2021mvw,DallaValleGarcia:2023xhh}. 

The ALPs and dark photons are widely used ``benchmark'' models proposed by the Physics Beyond Collider initiative~\cite{Beacham:2019nyx,Antel:2023hkf}. However, the implementation's flexibility makes it easy to add other models, such as the ALPs with the generic coupling pattern; the latter includes the class of models with non-minimal interactions of the type $\partial_{\mu}aH^{\dagger}i\overset{\leftrightarrow}{D^{\mu}}H$, inducing the $c_{q}$ couplings. Changing the $c_{q}$ pattern, we would change the representation of ALPs in terms of the $SU(3)$ generators. It may enhance or suppress the contribution of the mixing with the given meson, and, hence, has a large impact on the ALP phenomenology.

Below, we discuss it in detail.

\subsection{Incorporation in \texttt{PYTHIA8}}
\label{sec:implementation-pythia}
To simulate the production via fragmentation, we modified the source code. Consider the LLP having the mixing with the meson $m^{0}$. Specifically, we extended the method responsible for assigning particle IDs during fragmentation: each time when $m^{0}$ appears during the fragmentation, it is replaced with the LLP $X$ with a small seed probability of the order of $0.01$. Having simulated the production, we weight the event with the true production rate $|\Theta_{m^{0}X}|^{2}$ given by Eq.~\eqref{eq:theta-eff}. This recipe automatically allows us to account for the mass dependence of the LLP kinematics and the overall flux. 

As we already mentioned in Sec.~\ref{sec:challenges}, this implementation works only for the ground state mesons:
\begin{equation}
   m^{0} = \pi^{0},\eta,\eta',\rho^{0},\omega,\phi
\end{equation}
The higher excitations are not implemented in \texttt{PYTHIA8}; adding them would require tuning \texttt{PYTHIA8} setup to describe (unavailable or extremely hardly accessible) data on the production of such particles in high-energy proton collisions. Therefore, our results on the production yield in the mass range $m_{X}\gtrsim 1\text{ GeV}$, where the contributions from these resonances become important, are conservative.

Decays of mesons are implemented in the following way. For each model, we first consider a set of decay modes from Table~\ref{tab:decay-processes} and generate decays of mesons into LLPs at some fictitious branching ratio. Then, we weigh the events with the true model-dependent decay widths. Depending on the simplicity of the decay width as a function of the LLP's mass, we either provide an analytic expression for the branching ratio or its tabulated version.

Having simulated various production mechanisms, we merge the LLP entries from decays and fragmentation, reweight, and produce tabulated events.

\subsubsection{Experimental and simulation setups}
\label{sec:setup}

We consider four facilities housing most of the experiments relevant to searching for LLPs: Fermilab Beam Dump (further \texttt{FermilabBD}), SPS, Serpukhov (which housed the NuCal experiment~\cite{Blumlein:1990ay}), and LHC. The first three facilities are beam dump setups with the incoming proton beam energy, correspondingly, $E_{p} = 120, 400, 70\text{ GeV}$.

The strongest laboratory constraints and sensitivities at these facilities on the models of interest come from the experiments located in the forward direction~\cite{Beacham:2019nyx}. It is related to the fact that the solid angle distribution of light mesons is maximal in the domain of polar angles $\theta \lesssim \Lambda_{\text{QCD}}/p_{\text{forward}}$, where $p_{\text{forward}}$ is the characteristic energy of the forward particles (say, at the LHC, it is of the order of $1\text{ TeV}$); at higher angles, it typically quickly falls.\footnote{Depending on the target's thickness for the beam dump setups, a significant amount of mesons can be produced secondarily -- via cascade interactions from the primary collisions; similarly, we may expect a large flux of secondary $X$\!s. However, we expect them to increase the yield of $X$\!s in the forward direction only marginally. Indeed, secondary particles typically have a wide angular spread, and only a small fraction of them would fly in the forward direction relevant for most of the experiments that may probe their parameter space. See, e.g., Ref.~\cite{SHiP:2020vbd}, which discusses the contribution of dark photons from decays of secondary $\pi^{0},\eta,\eta'$.}

It is crucial to define the \texttt{PYTHIA8} setup used for simulating the events at various facilities. It includes the PDF set, the type of processes used to simulate the interactions, and a set of various parameters that may be tuned to accurately match the predictions with the experimental data. As we are interested in the forward direction, for the LHC, we will utilize the \texttt{forward tune} from~\cite{Fieg:2023kld}. We will use this fit in Sec.~\ref{sec:case-studies} to evaluate the uncertainties in the flux of $X$\!s.

For the beam dump setups, no careful tunes exist in the literature. However, Ref.~\cite{Dobrich:2019dxc} considered a setup to describe the data from various proton beam energies. In particular, for SPS energies, it agrees within 20\% to the full kinematic data~\cite{NA62:2023qyn}. We do not expect the uncertainties to inflate for the beam dumps, but a dedicated study clarifying this point will be conducted once such tunes are made. 

In particular, the amounts of $\pi^{0}$ and $\rho^{0}$ mesons generated by \texttt{PYTHIA8} with the specified setups are 
{\small
\begin{multline}
(P_{\text{prod},\pi^{0}},P_{\text{prod},\rho^{0}}) =\begin{cases}
        (2.87,0.37), \quad \text{FermilabBD} \\ (4.14,0.55), \quad \text{SPS}\\ (32.03,4.51), \quad \text{LHC} \\ 
    \end{cases}
    \label{eq:yield-mesons}
\end{multline}
}
\subsection{Incorporation in \texttt{SensCalc}}
\label{sec:implementation-senscalc}

\texttt{SensCalc}~\cite{Ovchynnikov:2023cry} is a \texttt{Wolfram}-based~\cite{Mathematica} tool that calculates the number of events with various LLPs at different experiments. \texttt{SensCalc} simulates the production and propagation of LLPs toward the decay volume, incorporating in particular the effects of the LLP lifetime on the event distribution. The code includes various experiments located at various facilities, accounts for the geometry of the decay volume and the detector of the given experiment, and the presence of a magnetic field in the spectrometer, bending the trajectories of decay products. Finally, it incorporates various LLPs, including Heavy Neutral Leptons, dark matter models, the LLPs considered in this study, and other cases. 

Users may select the LLP production and decay channels, event signatures, and selection criteria on the decay products used to compute the constraints and sensitivities. This way, the tool is very versatile and flexible.

Initially, the tool only computed the number of events without producing a detailed event record. However, recently, a \texttt{SensCalc}-based Monte Carlo sampler of the events has been added as well~\cite{Kyselov:2024dmi}. Although the primary goal of the tool is to compute the exclusion region for the decaying LLPs, other signatures can be studied, such as dark matter scattering or events with simultaneous decays of two LLPs. A public version of this updated code will be released soon.

\subsubsection{Incorporating production modes via mixing}
The step preceding the calculation of the number of events depends on the production mode.  It either generates the angle-energy distribution of LLPs in-flight or uses the pre-generated tabulated distribution provided by an external source (e.g., \texttt{PYTHIA8}). \footnote{This step is LLP lifetime-independent, being determined solely by the LLP mass.} The first approach is convenient because it is flexible and can account for the complicated matrix elements of the LLP production mechanisms without the need to implement them in event generators. The second approach is unavoidable, however, in the case of LLP production in deep-inelastic collisions, where the details of the Drell-Yan process and fragmentation must be simulated explicitly.

Previously, the production in the fragmentation chain for the vector mediators and collective production from the three processes in Fig.~\ref{fig:production-channels} for the ALPs has been implemented via Eq.~\eqref{eq:state-of-the-art-alps}, following the state-of-the-art description of these modes; the corresponding production mechanisms were called \texttt{Mixing}. 

In the recent version, we have updated these channels for each of the models:
\begin{itemize}
    \item[--] Decays of mesons: we use the meson tabulated angle-energy distributions pre-generated by \texttt{PYTHIA8}, sample their kinematics inside \texttt{SensCalc}, and then simulate their decays into LLPs using the decay matrix elements. 
    \item[--] Proton bremsstrahlung: we did not change the available description for the dark photons, but implemented this mechanism for the ALPs from scratch. This is done in the same way as it has been implemented for the other particles -- considering the three choices of the proton virtuality scale $\Lambda_{p} = 0.5,1,2\text{ GeV}$, which allows estimating the impact of uncertainty in production (see also Ref.~\cite{Kyselov:2024dmi}). The relevant production modes are termed \texttt{Bremsstrahlung-AP} -- named after the initials of the authors who first employed this approach for the first time~\cite{Altarelli:1977zs} (see also Ref.~\cite{Boiarska:2019jym}) -- and \texttt{Bremsstrahlung-FR}, which refers to the method introduced in Ref.~\cite{Foroughi-Abari:2024xlj} that builds upon the AP approach for dark photons by eliminating the unphysical longitudinal divergence.
    \item[--] Parton fragmentation: we pre-generated tabulated angle-energy distributions of LLPs using our \texttt{PYTHIA8} modification. The production mechanism is called \texttt{Fragmentation}. The old \texttt{Mixing} description has been renamed to \texttt{Old-Mixing}.
\end{itemize}

Users may switch between the state-of-the-art description of the production at the stage of the computation of sensitivities of different experiments. 

A side update in \texttt{SensCalc} includes adding theoretical uncertainties in the decay widths of Higgs-like scalars. Two descriptions are available: following Ref.~\cite{Winkler:2018qyg} and Ref.~\cite{Blackstone:2024ouf}, where, in the latter case, we added the variation of the width with the internal parameters of the calculation method. Combining the uncertainties in production and decay, it is possible to derive the total effect of the uncertainty in the LLP phenomenology on constraints and sensitivities.

\section{Toy study: simple mixing pattern}
\label{sec:toy}

In this section, we utilize the implementation and illustrate the impact of LLP mass on the production via mixing. We assume artificial scenarios of LLPs mixing with one meson and consider the yield and kinematic distributions. For these setups, we calculate the branching ratios of decays of different mesons into $X$\!s from Table~\ref{tab:decay-processes} just by rescaling the SM branching ratios by appropriate kinematic factors, similar to the one in Eq.~\eqref{eq:br-DP}.

\begin{figure}[t!]
    \centering
    \includegraphics[width=0.45\textwidth]{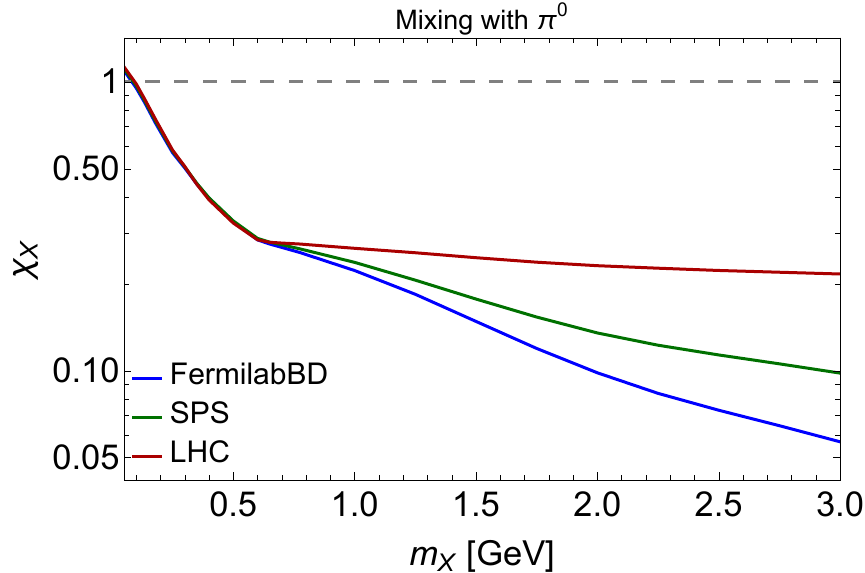}
    \caption{The yield~\eqref{eq:normalized-yield} of particles $X$ mixing with $\pi^{0}$ produced per proton collision via the fragmentation and decays of secondary mesons, for various facilities -- Fermilab Beam Dump (\texttt{FermilabBD}), SPS, and LHC. The yield is normalized by the amount of the corresponding mesons, which we report in Eq.~\eqref{eq:yield-mesons}. The quick decrease of the yield in the mass domain $m_{X}\lesssim m_{\rho}$ is caused by closing secondary decays of heavier mesons ($\rho^{0},\eta,\eta',\omega,K_{S}$) into $X$ after reaching the kinematic threshold. Here and below, we use \texttt{PYTHIA8} tunes are described in Sec.~\ref{sec:setup}.}
    \label{fig:production-magnitudes}
\end{figure}

Let us start with the $X$ yield, which we define by
\begin{equation}
    P^{m^{0}}_{\text{prod,X}} = \frac{\sigma^{m^{0}}_{pp\to X+\text{other}}}{\sigma_{pN\to \text{all}}}
    \label{eq:production-probability}
\end{equation}
Here, $\sigma^{m^{0}}_{pp\to X}$ is the production cross-section, scaling as 
$\sigma^{m^{0}}_{pp\to X+\text{other}} = f(m_{X}) |\theta_{Xm^{0}}|^{2}$, with $f(m_{X})$ being the function accounting for the LLP mass dependence of the flux modulus the squared mixing angle $|\theta_{Xm^{0}}|^{2}$. Next, $\sigma_{pN\to\text{all}}$ is the proton-nucleon cross-section at the given setup. For the LHC, it is just the proton-proton cross-section, whereas for the beam dump facilities, it is the averaged proton-nucleon cross-section. 

\begin{figure}[h!]
    \centering
    \includegraphics[width=0.45\textwidth]{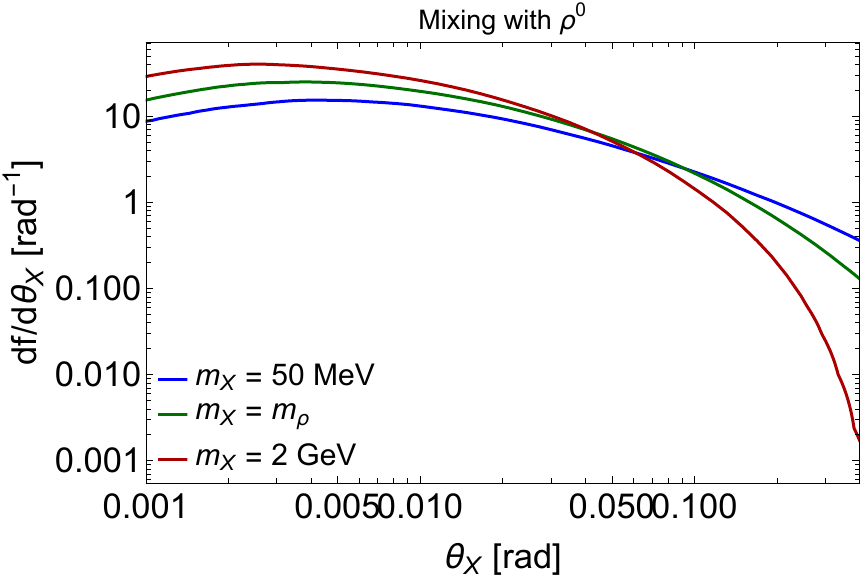}\\ \includegraphics[width=0.45\textwidth]{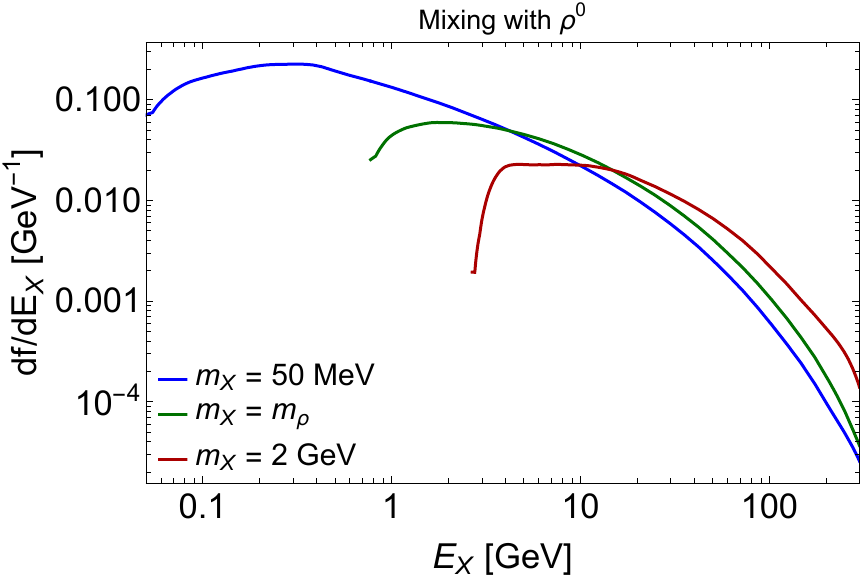}
    \caption{Polar angle and energy distributions of the particles $X$ mixing with $\rho^{0}$ meson, for various choices of the $X$ mass $m_{X} = 50\text{ MeV}, m_{\rho}, 2\text{ GeV}$. The fragmentation production channel and the SPS facility are considered. The distributions are normalized by one.}
    \label{fig:distributions}
\end{figure}

To study the impact of the $X$'s mass on the flux, we introduce the normalized probabilities 
\begin{equation}
\chi^{m^{0}}_{X} \equiv \frac{P^{m^{0}}_{\text{prod,X}}}{|\theta_{Xm^{0}}|^{2}P_{\text{prod},m^{0}}}
\label{eq:normalized-yield}
\end{equation}
We first checked the baseline consistency of our framework by recovering $\chi^{m^{0}}_{X}(m_{X} = m_{0})\approx 1\cdot |\theta_{Xm^{0}}|^{2}$. Precisely, the prefactor in front of $|\theta_{Xm^{0}}|^{2}$ is one in all the cases except for the mixing with $\pi^{0}$. For the latter, we got 0.88 for SPS, 0.89 for the LHC, and 0.86 for BD@Fermilab. The lack of the events corresponds to decays of the baryon resonances $\Lambda, \Delta^{0}, \Delta^{++}$, see Appendix~\ref{app:pion-production-from-decays}. While they may be described analogically to the implemented processes, we will not consider them, as the phase space available for $X$ is very limited. 

\begin{figure*}[t!]
    \centering
    \includegraphics[width=0.5\textwidth]{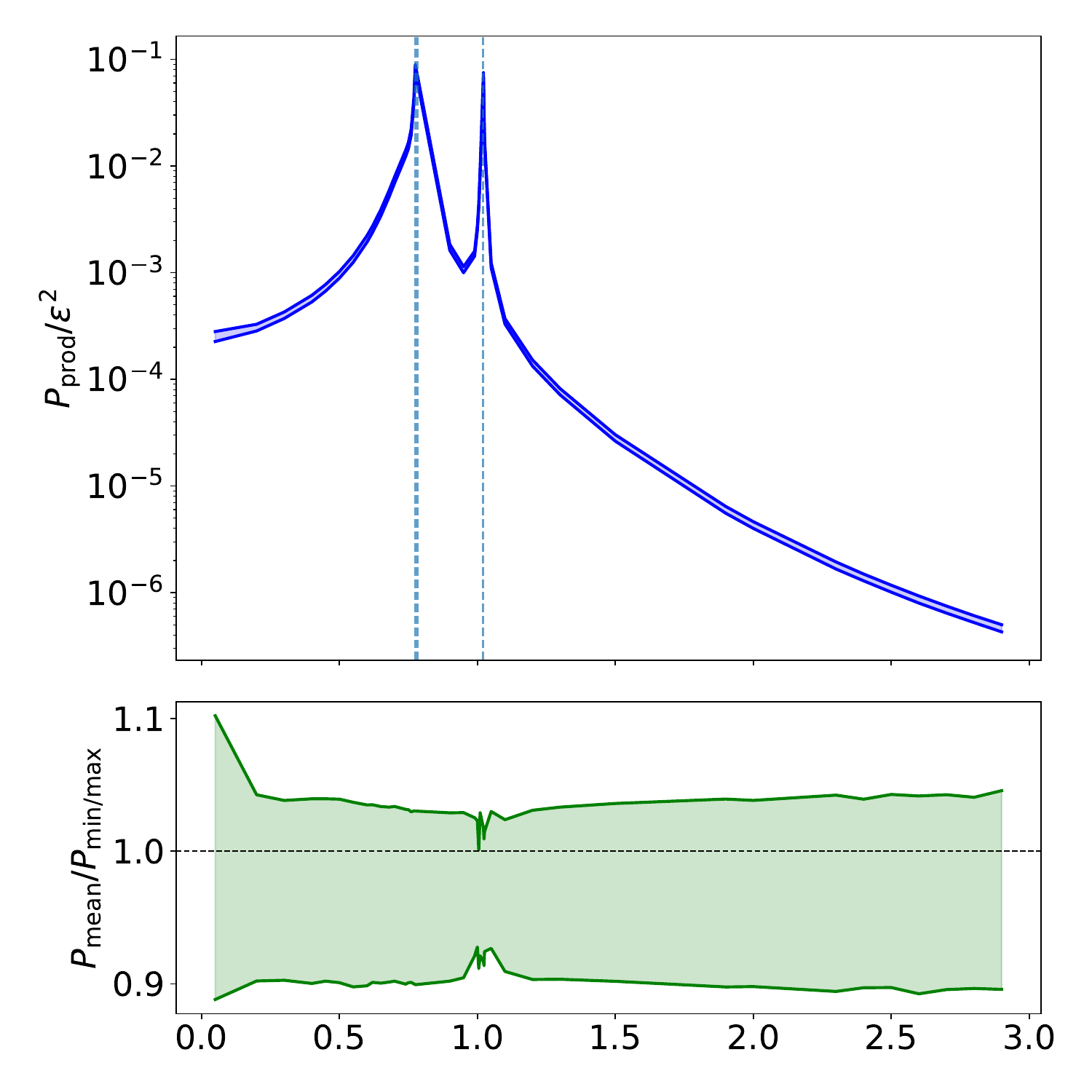}~\includegraphics[width=0.5\textwidth]{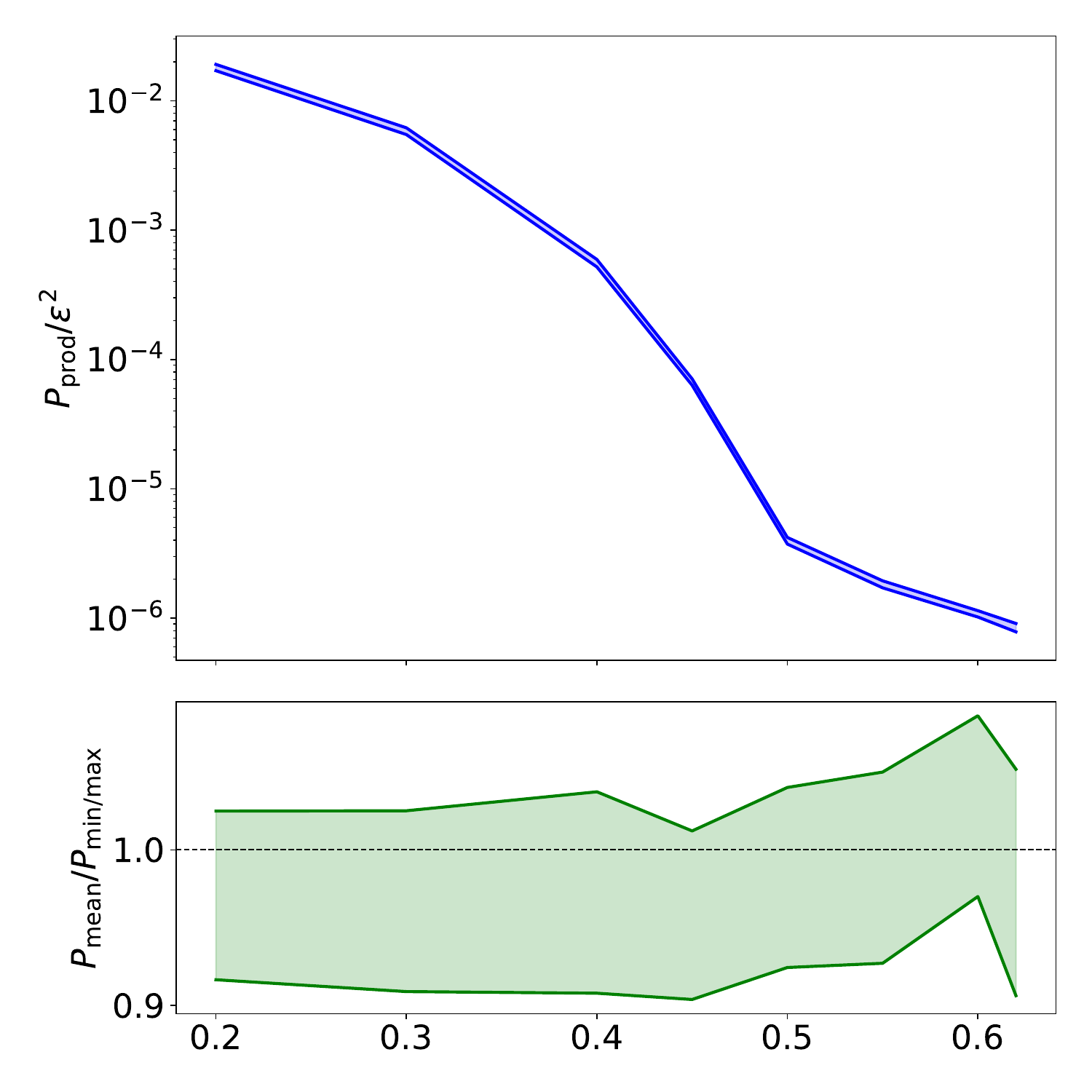}
    \caption{The yield of dark photon particles per squared coupling $\epsilon^{2}$ (see Eq.~\eqref{eq:production-probability}) produced at the LHC within the angular domain $\theta < 1\text{ mrad}$. It has been calculated using the far-forward tune from~\cite{Fieg:2023kld}. The left panel shows the production in the fragmentation chain, whereas the right one shows the contribution of decays of $\pi^{0},\eta,\eta'$ mesons. The uncertainty band (in blue) is obtained by varying the parameters within the ranges specified in Table~I of the paper. The variation of the predictions with the fit parameters is indicated as the blue band top plot, and the relative errors are represented as a green domain in the subplot.}
    \label{fig:uncertainties}
\end{figure*}

We show the behavior of the quantity~\eqref{eq:normalized-yield} for the mixing with pions in Fig.~\ref{fig:production-magnitudes}. The generic pattern is that for masses $m_{X}<m_{m^{0}}$, $\chi_{X}$ is slightly larger than $1$, while at larger masses, it falls down. In the mass range $m_{X}\lesssim 600\text{ MeV}$, the behavior is very similar for all the facilities, $\chi_{X}^{\pi^{0}}$ drops by a factor of $1/3$. It is because, in all cases we considered, around $70\%$ of the produced $\pi^{0}$\!s originate from decays of heavier mesons (see Appendix~\ref{app:pion-production-from-decays}). By increasing the $X$ mass, we quickly decrease the yield of such mesons. On the other hand, the amount of $X$s produced by the fragmentation changes much slower in this mass range, so the evolution of $\chi$ is dominated by decays. At larger masses, $\chi_{X}$ is controlled solely by the fragmentation channel; it continues falling but differently for the three facilities -- the lower the collision energy is, the smaller the yield is. Overall, compared to the naive treatment~\eqref{eq:state-of-the-art-alps}, the drop may be as large as a factor of 20. 

Next, let us analyze $X$'s kinematics. In Fig.~\ref{fig:distributions}, we show the polar angle and energy distributions of $X$ for the mixing with $\rho^{0}$ produced by the fragmentation at the SPS facility. We consider the two cases $m_{X} = 50\text{ MeV}$ and $m_{X} = 3\text{ GeV}$. The common pattern is that with the mass decrease, the polar angle distribution becomes narrower while the energy distribution gets shifted to higher values. These results are what we expected naively for heavier particles. 

\section{Case studies}
\label{sec:case-studies}

In this section, we consider the realistic models discussed in Sec.~\ref{sec:state-of-the-art}: dark photons, mediators coupled to the baryon current, and axion-like particles with various coupling patterns.

\subsection{Production in fragmentation and decays of mesons: uncertainties}
\label{sec:uncertainty-our-approach}

Let us first estimate the uncertainty for dark photon production in decays of mesons and fragmentation. We will consider the LHC case, for which we utilize the forward tune from Ref.~\cite{Fieg:2023kld} (remind a discussion in Sec.~\ref{sec:state-of-the-art}). To assess uncertainty, we analyze how the yield of new particles changes when varying the parameters $\sigma_{\text{soft}}$ and $\sigma_{\text{hard}}$ of the tune within the ranges specified in Table~I of that paper. Since this tune is designed for the pseudorapidity range $\eta \geq 9$, we shall also constrain the angular range for the particles produced to $\theta < 1\text{ mrad}$.

The uncertainty band in the yields of dark photons produced in a single proton-proton collision through the fragmentation and decays of the meson channels are shown in Fig.~\ref{fig:uncertainties}.\footnote{A similar uncertainty is observed for the mediators coupled to the baryon current.} The uncertainty is mostly within 20\%. Our results for the dark photons produced by decays are consistent with Fig.~4 from Ref.~\cite{Fieg:2023kld}.\footnote{The uncertainty can be recovered from the right panel of this plot by taking into account that at the lower bound of the sensitivity the number of events with dark photons scales as $\epsilon^{4}$.}

A very similar uncertainty at the LHC is the case for the other models considered in this study. However, as we have already discussed (see Sec.~\ref{sec:implementation-pythia}), there are no detailed studies on the tunes for the other facilities. Nevertheless, we do not expect dramatic deviation of the uncertainties in the latter case from the LHC one.

\subsection{Vector mediators}
\label{sec:case-study-vectors} 

\begin{figure*}[t!]
    \centering
    \includegraphics[width=0.5\textwidth]{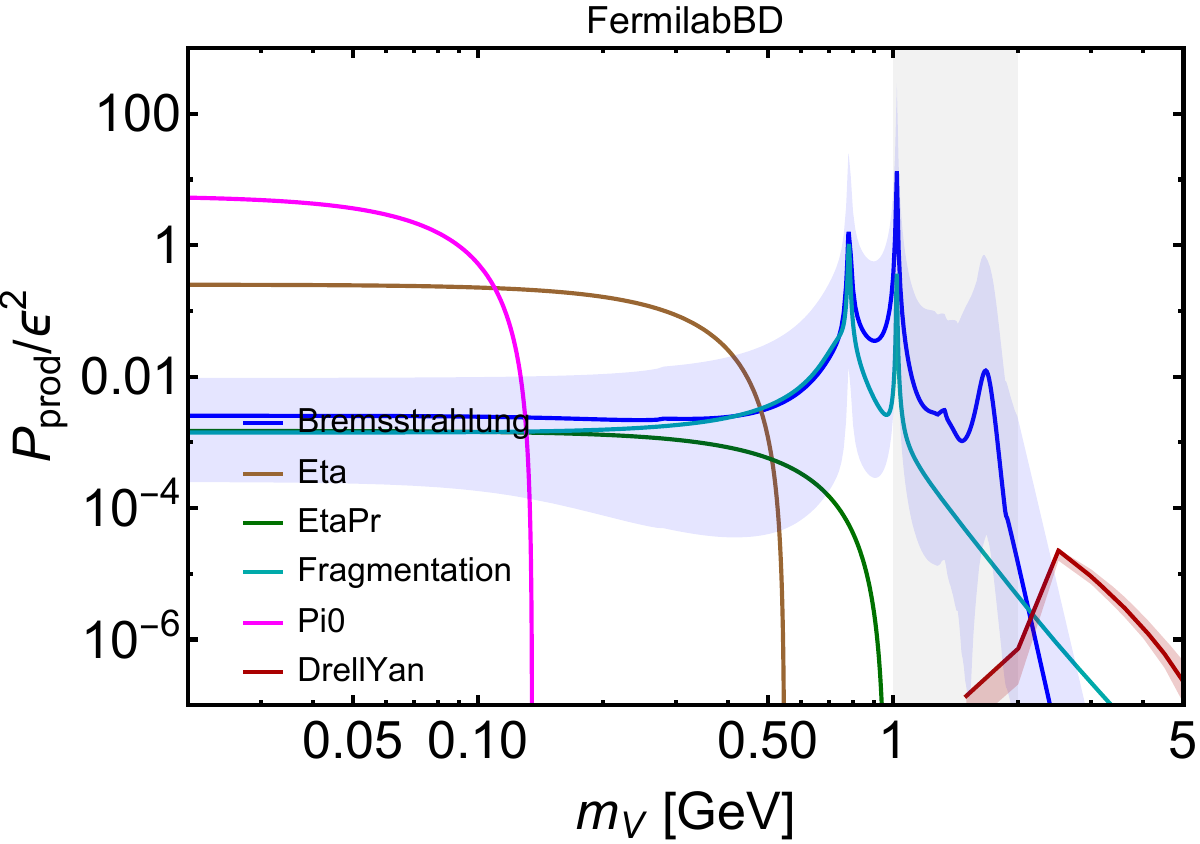}~\includegraphics[width=0.5\textwidth]{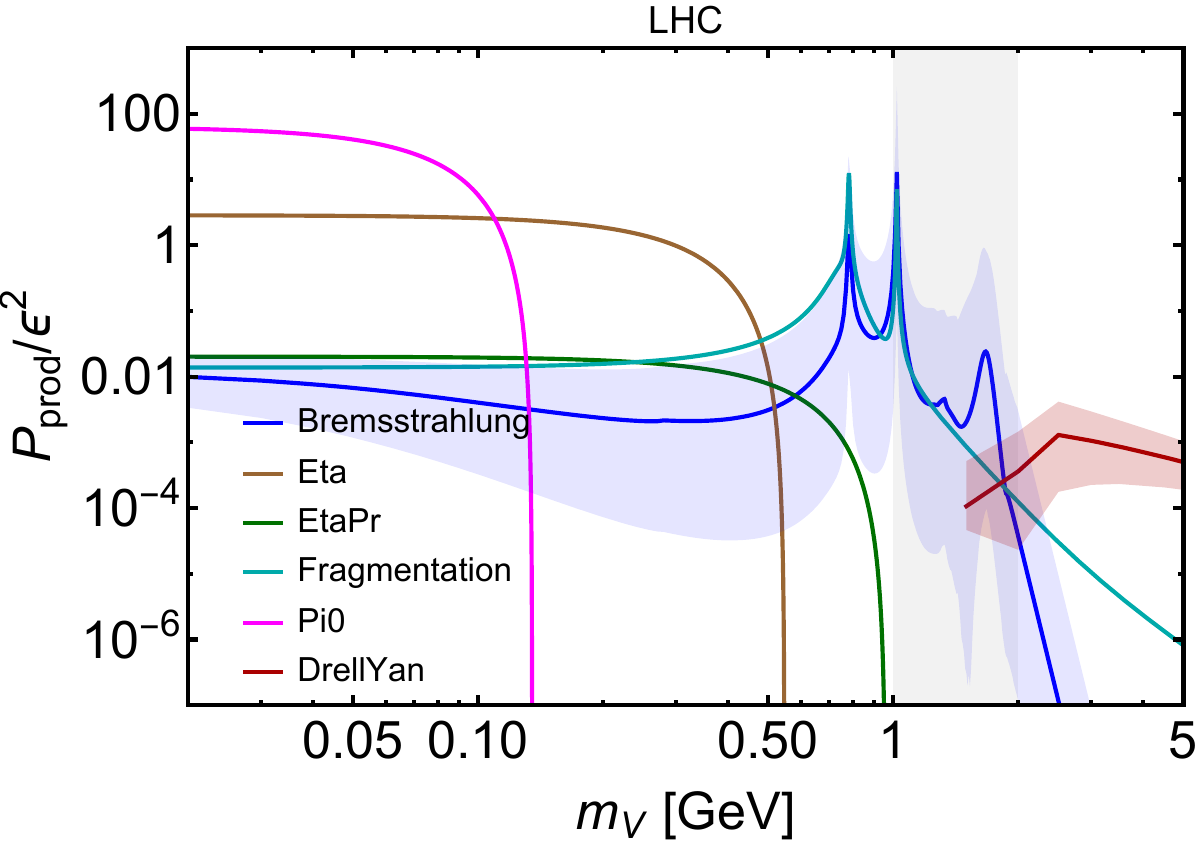}
   \\
    \includegraphics[width=0.5\linewidth]{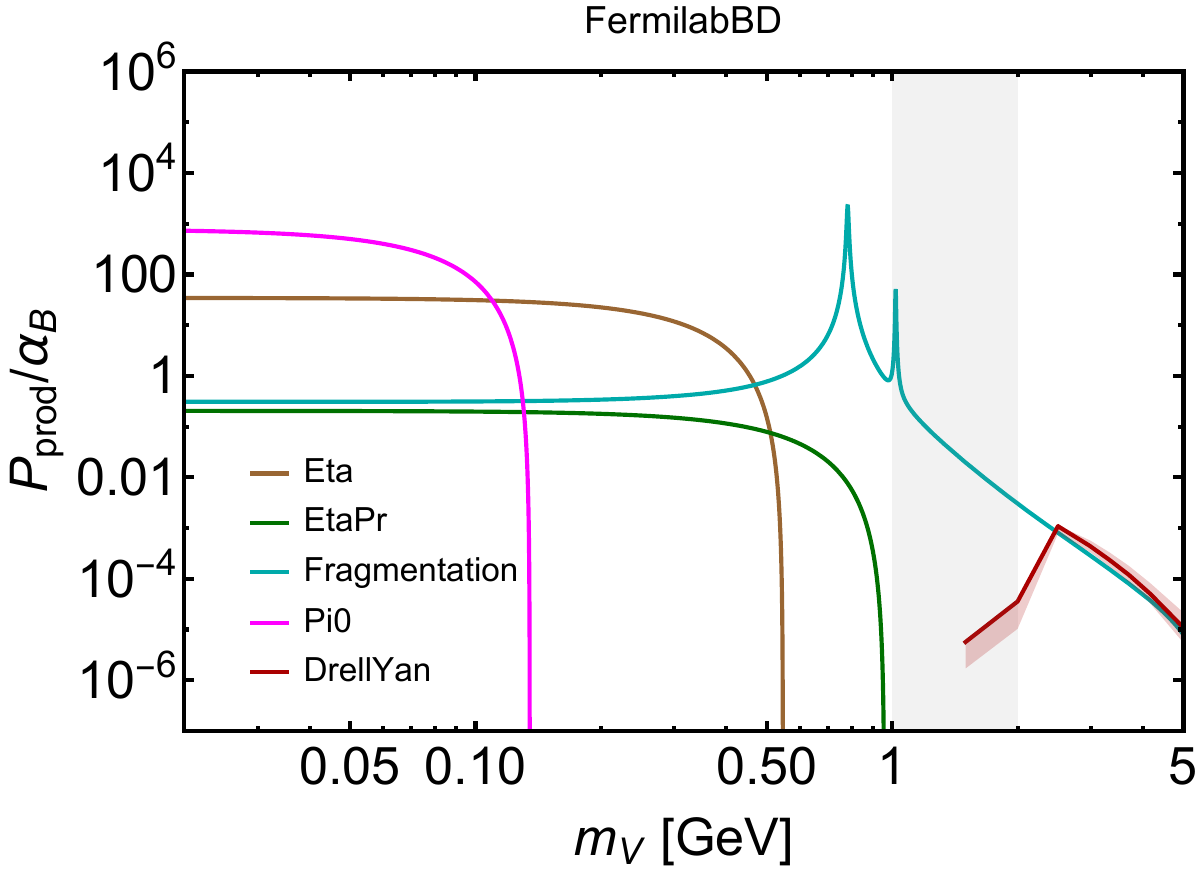}~\includegraphics[width=0.5\linewidth]{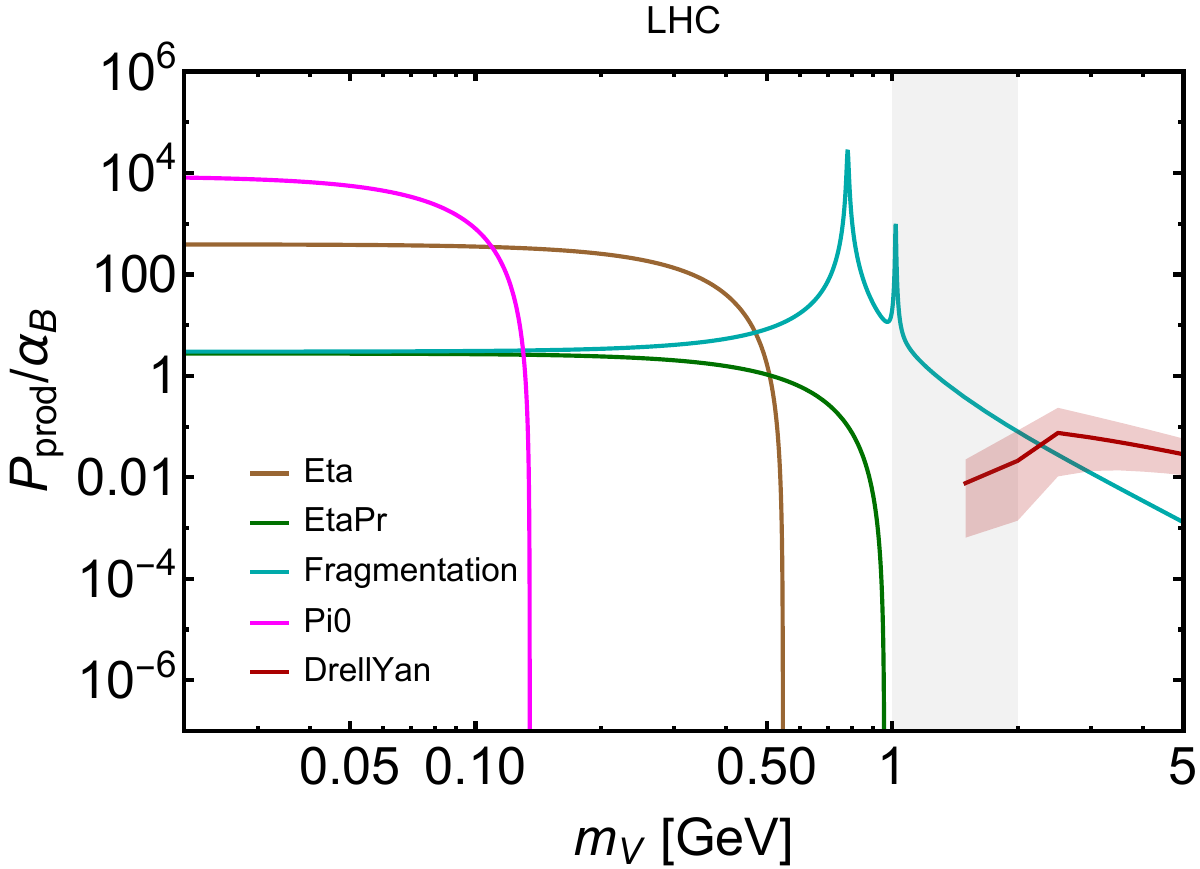}
    \caption{Overall production probabilities of vector mediators (Eq.~\eqref{eq:production-probability}) from various processes at FermilabBD (left plots) and LHC (right plots), for dark photons (top panel) and the mediators coupled to the baryon current (bottom panel). The figures have been obtained using \texttt{SensCalc} package~\cite{Ovchynnikov:2023cry}, where we incorporated the production mechanism via fragmentation and decay of mesons as obtained in this work. See Ref.~\cite{Kyselov:2024dmi} for the description of the uncertainties in the bremsstrahlung and Drell-Yan production modes. We do not show the uncertainties in the channels of fragmentation and decays of mesons, provided that they are quite small. In addition, we do not show the bremsstrahlung flux for the mediators coupled to the baryon current, given the immature status of the bremsstrahlung description for them. The gray band in the mass range $1\text{ GeV}<m_{V}<2\text{ GeV}$ shows the domain where the production may be sizeably affected by the mixing with higher vector resonances not included in the calculations (see Sec.~\ref{sec:challenges}).}
    \label{fig:production-dark-photon}
\end{figure*}

Now, let us compare different production modes, starting with vector mediators. For dark photons $V$, apart from decays of mesons, proton bremsstrahlung, and fragmentation, the other important mechanism is the quark fusion, also known as the Drell-Yan process~\cite{Ovchynnikov:2023cry,Kyselov:2024dmi}. The dominant meson decay modes are $\pi^{0}/\eta/\eta'\to \gamma V$. 

The production palette of the mediators coupled to the baryon current closely resembles the dark photon case, with two important differences. First, due to the intrinsic properties of the baryon current, these mediators only mix with $\omega,\phi$ families, not with $\rho$. Therefore, the production is enhanced only at the narrow vicinity of $\omega,\phi$ masses. The second difference is in the description of the proton bremsstrahlung. As we mentioned in Sec.~\ref{sec:mixing-definition}, there are no reliable calculations of the proton elastic form factor for the mediators coupled to the baryon current. Therefore, we do not include this production channel.

\begin{figure*}[t!]
    \centering
    \includegraphics[width=0.5\linewidth]{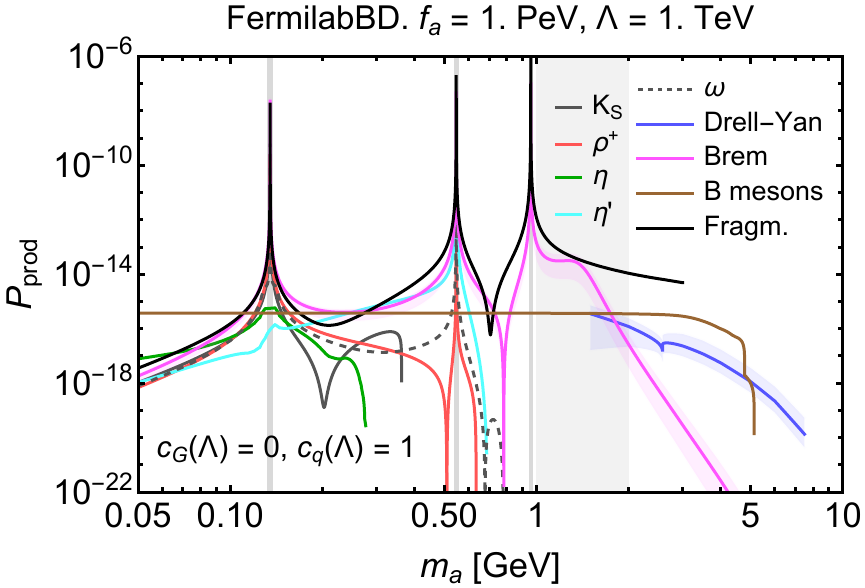}~\includegraphics[width=0.5\linewidth]{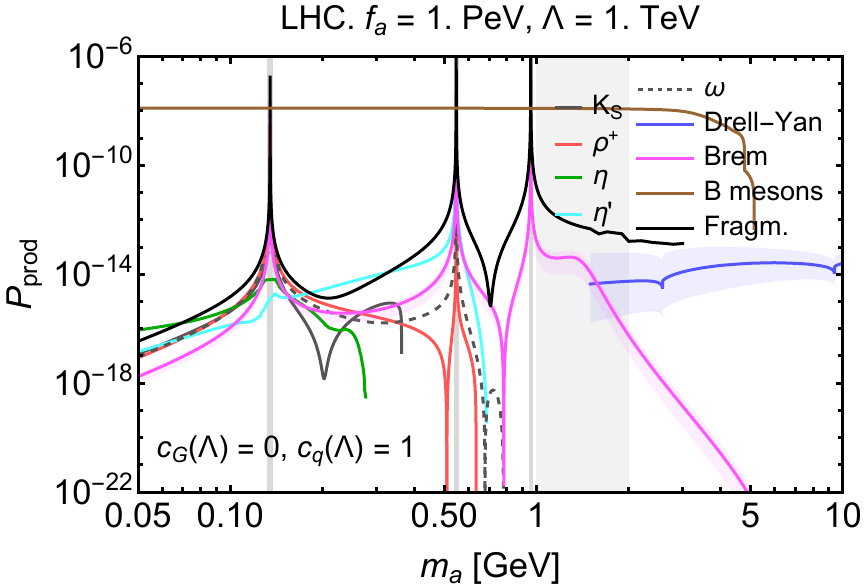}
    \\     \includegraphics[width=0.5\linewidth]{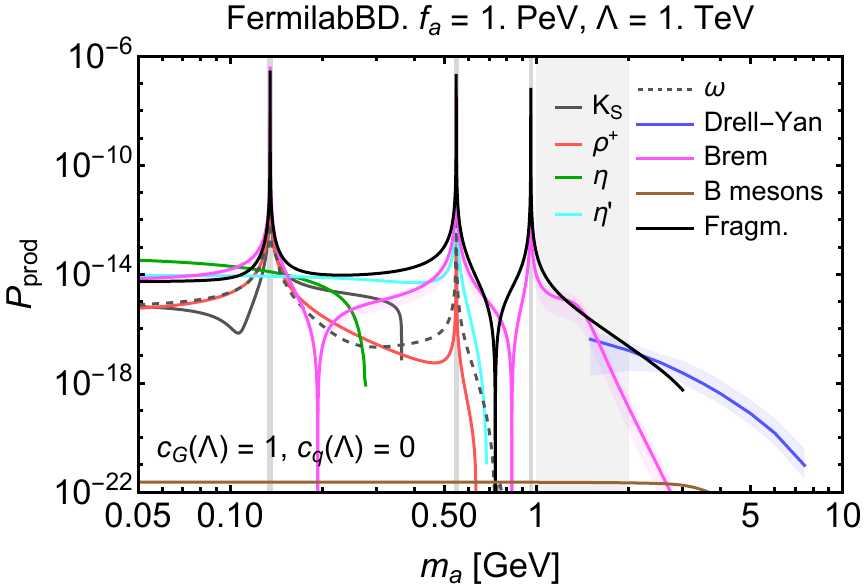}~\includegraphics[width=0.5\linewidth]{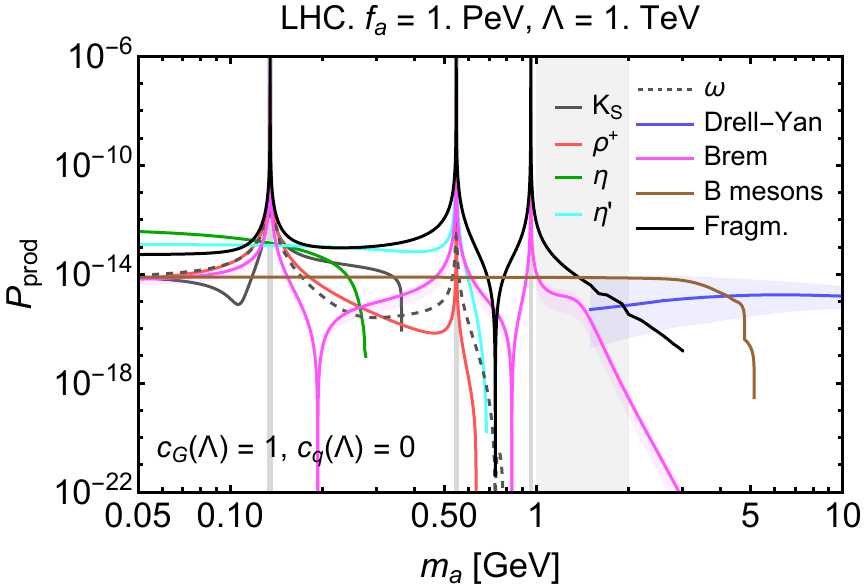}
    \caption{Production probabilities of the axion-like particles at FermilabBD and LHC facilities. The ALP models are defined by the Lagrangian~\eqref{eq:lagrangian-alp} at the scale $\Lambda = 1\text{ TeV}$, corresponding to the traditional choice in the literature~\cite{Beacham:2019nyx,Antel:2023hkf,DallaValleGarcia:2023xhh}. Top panel: ALPs with the universal coupling to quarks ($c_{q} = 1, c_{G} = 0$ in Eq.~\eqref{eq:lagrangian-alp}). Bottom panel: ALPs coupled to gluons ($c_{q} = 0, c_{G} = 1$). The peaks are explained by the resonant contribution of the mixing with $m^{0} = \pi^{0},\eta,\eta'$ mesons, whereas the dips are caused by the interference between these contributions and the direct interaction terms. The gray bands show the vicinities of the masses of $m^{0}$ where the description of phenomenology based on the mixing breaks down and the domain $m_{a}\gtrsim 1\text{ GeV}$ where the description of the ALP phenomenology is incomplete due to lack of knowledge about excited pseudoscalar mesons (remind Sec.~\ref{sec:challenges}).}
    \label{fig:production-ALP}
\end{figure*}

Fig.~\ref{fig:production-dark-photon} shows the production probabilities of vector mediators at the LHC and FermilabBD. The common pattern is that the production from mesons dominates in the mass range $m_{V}\lesssim m_{\eta}$. At larger masses, direct mechanisms become the only possible mode. As we have discussed in Sec.~\ref{sec:challenges}, the proton bremsstrahlung has a large uncertainty caused by the parameterization of the proton elastic form factor in the proton-proton-$V$ vertex and the unknown behavior of the proton virtuality~\cite{Foroughi-Abari:2021zbm,Foroughi-Abari:2024xlj}. We have computed the flux of vectors produced in the Drell-Yan process by simulating the leading-order processes of the gluon and quark fusion following~\cite{Ovchynnikov:2023cry,Kyselov:2024dmi}. The models of vector mediators have been implemented with the help of \texttt{FeynRules}~\cite{Alloul:2013bka,Christensen:2008py}. For the parton distribution function, we use \texttt{NNPDF 3.1 NNLO}. The uncertainty of the Drell-Yan production has been estimated by varying the factorization and renormalization scales by a factor of two around their central values.

The production via fragmentation competes with the production via the bremsstrahlung process in the whole mass range. However, in the domain $m_{V}\gtrsim 1\text{ GeV}$, it is at the lower part of the uncertainty bound of bremsstrahlung. It is because, as we discussed in Sec.~\ref{sec:implementation}, the fragmentation does not include the contribution from higher resonances like $\rho^{0}(1420), \rho^{0}(1700), \dots$, which are expected to dominate the production at high masses. On the other hand, the bremsstrahlung description includes them via contributions to the elastic form factor. 

\subsection{ALPs}
\label{sec:case-study-alp}

Now, let us proceed to the ALPs. Overall, the production mechanisms are very similar to the dark photon case, with one addition -- there is an important mode of decay of $B$ mesons and kaons. They emerge even from the Lagrangian~\eqref{eq:lagrangian-alp} with flavor-diagonal couplings to quarks. The reason is the renormalization group flow from the scale $\Lambda$, at which the Lagrangian is defined, down to the scale $Q \simeq m_{a}$ of the ALP production~\cite{Bauer:2021wjo}. The latter induces flavor-changing neutral current transition operators such as $b\to s$ and $s\to d$ (decays $K\to a + \pi$).

\begin{figure}[t!]
    \centering
    \includegraphics[width=\linewidth]{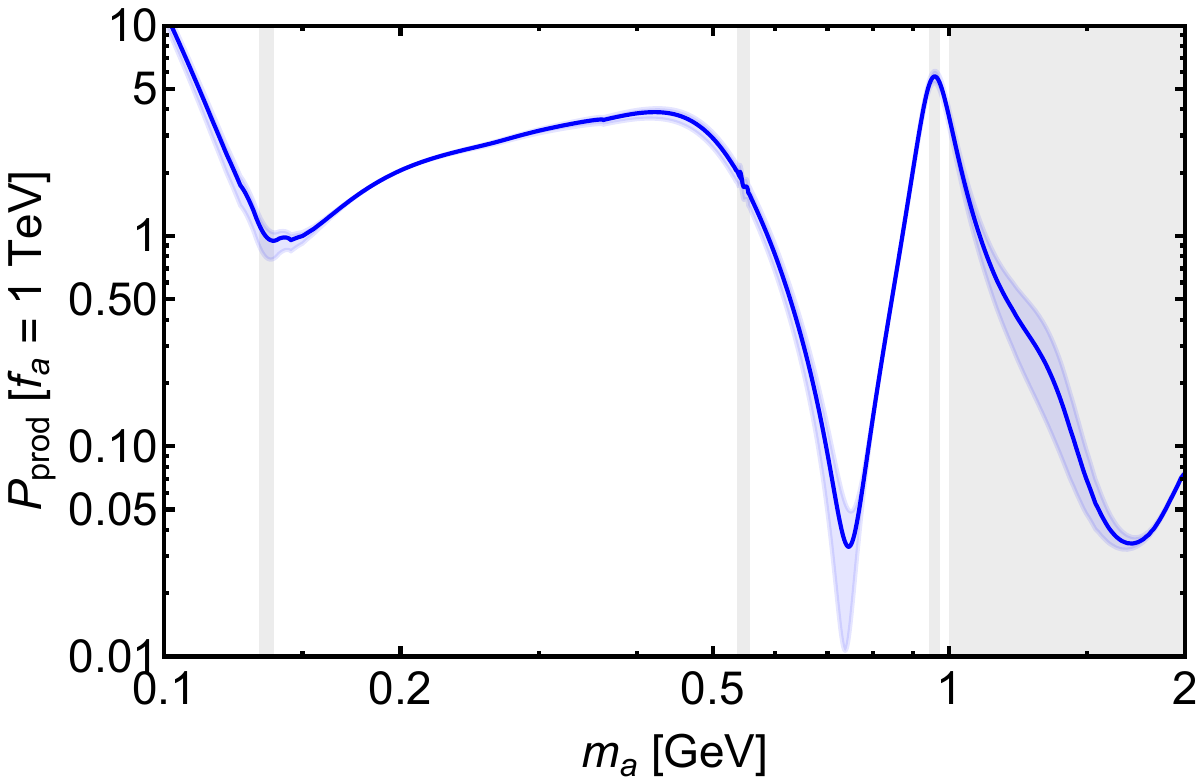}
    \caption{The ratio of the total ALP production probability at FermilabBD as obtained in this study to the one derived using Eq.~\eqref{eq:state-of-the-art-alps}, commonly adopted in the previous studies~\cite{Jerhot:2022chi,Antel:2023hkf,Ovchynnikov:2023cry}. The model with the ALPs coupled solely to gluons is considered.}
    \label{fig:alp-production-old-vs-new}
\end{figure}

Provided the universal coupling of ALPs to quarks, $c_{q} = 1$, the decay of $B$ mesons can dominate the production; the decays under consideration are $B\to X_{s}+a$, where $X_{s} = K,K^{*},K_{1},\dots,$ \textit{etc}. There are a few caveats here~\cite{DallaValleGarcia:2023xhh}. First, the scale $\Lambda$ defining the effective Lagrangian~\eqref{eq:lagrangian-alp} with $c_{t} \neq 0$ must be $\Lambda \gtrsim 1\text{ TeV}$; otherwise, the induced $b\to s$ coupling would not be sizeable, and $B$ decays into ALPs become suppressed.\footnote{We should acknowledge, however, that this statement depends on the ALP UV completion details, since there may be in principle operators that provide unsuppressed yield, as well as the presence of the other ALP couplings (such as the $aWW$ interaction), which induce this production as well, independently of the $c_{t}$ contribution.} Second, the center-of-mass collision energy $\sqrt{s}$ at the given facility must be far enough from the di-B threshold $2m_{B}$; otherwise, the yield of $B$ mesons would be a priori suppressed. 

The facilities falling into the ``low energy'' category are FermilabBD and Serpukhov. Finally, the angular position of the experiment is also important. The solid angle flux of ALPs produced directly in the proton-proton collisions is peaked in the far-forward direction, whereas the $B$ flux is more isotropic. Therefore, for the experiments placed in the far-forward direction, such as BEBC~\cite{BEBCWA66:1986err}, the direct production flux gets a relative enhancement.

Fig.~\ref{fig:production-ALP} shows the production pattern for the two ALP setups: those universally coupled to fermions or coupled to gluons (recall Sec.~\ref{sec:implementation}). We take the description of the ALP production in the proton bremsstrahlung from Ref.~\cite{Ovchynnikov:2025gpx} (see also Ref.~\cite{Blinov:2021say}). Similarly to the vector mediators case, there are sizeable uncertainties in the proton bremsstrahlung and Drell-Yan process. In addition, we do not include the contribution of the mixing with heavier mesons in the production by fragmentation (see Sec.~\ref{sec:challenges}). 

This figure illustrates the relative unimportance of $B$ meson decay to fragmentation in the GeV mass range due to the low collision energy.

We conclude our discussion by comparing the flux of ALPs as obtained in this work by combining the proton bremsstrahlung, decays of mesons, and fragmentation, and the flux calculated using the integrated version of the simplified approach~\eqref{eq:state-of-the-art-alps}. The comparison is shown in Fig.~\ref{fig:alp-production-old-vs-new}. Depending on the mass, the difference can be 1-2 orders of magnitude; apart from that, the ALP angle-energy distribution differs significantly from the naive result~\eqref{eq:state-of-the-art-alps}.

\section{Conclusions}
\label{sec:conclusions}

Recently approved lifetime frontier experiments will probe previously unexplored parameter space for new physics in the GeV mass range, with long-lived particles (LLPs) representing a key focus. In view of this, a comprehensive understanding of their phenomenology, particularly how they can be produced directly in high-energy proton-proton collisions, is essential.

In this work, we studied the class of models in which LLPs mix with neutral mesons; relevant examples include Higgs-like scalars, axion-like particles (ALPs) with various coupling patterns, and vector mediators such as dark photons or vectors coupled to the baryon current (see Sec.~\ref{sec:mixing-definition}). This mixing contributes to multiple production channels, including meson decays, fragmentation, and proton bremsstrahlung (Fig.~\ref{fig:production-channels}). However, current descriptions of these mechanisms (Sec.~\ref{sec:state-of-the-art}) rely on approximations and exhibit inconsistencies. In particular, a common approach that expresses the LLP flux as the meson flux multiplied by the squared mixing angle fails to account for the LLP mass dependence and, in some cases (e.g., for ALPs), introduces unphysical parameter dependencies. These limitations underscore the need for a more refined treatment.

In this paper, we performed a systematic study of the production mechanisms induced by meson mixing, introducing our framework in Sec.~\ref{sec:approach}. We highlighted the unavoidable uncertainties arising from the poorly known properties of mesonic excitations with masses $M>1\text{ GeV}$ (Sec.~\ref{sec:challenges}). These uncertainties severely influence the fragmentation and bremsstrahlung production modes. Then, we have implemented these production channels for several benchmark LLP models in both \texttt{PYTHIA8} and \texttt{SensCalc} (Sec.~\ref{sec:implementation}) tools. Our implementation is publicly available, and its modular structure allows for model- and experiment-agnostic usage.

Next, in Sec.~\ref{sec:toy}, we considered a simplified scenario where the LLP mixes with a single meson to illustrate how the LLP flux and kinematics vary with its mass, depending on the experimental facility. In Sec.~\ref{sec:case-studies}, we extended our analysis to realistic models, estimated the uncertainty in the production flux from fragmentation and meson decays (Sec.~\ref{sec:uncertainty-our-approach}), and compared the contributions from various production channels on a case-by-case basis.

Overall, our predictions for the LLP flux differ by 1-2 orders of magnitude from previous results (Figs.~\ref{fig:production-magnitudes} and \ref{fig:alp-production-old-vs-new}). The LLP kinematics also deviate substantially, especially for non-negligible LLP masses (Fig.~\ref{fig:distributions}). Along with existing uncertainties in LLP decay descriptions~\cite{Ovchynnikov:2025gpx,Blackstone:2024ouf}, these production uncertainties pose notable challenges for mapping the parameter space of GeV-scale new physics particles. Advancing our theoretical and experimental understanding of heavy meson properties is crucial for reducing these uncertainties.

\section*{Acknowledgments}

The authors thank Torbjörn Sjöstrand, Peter Skands, Tim Cohen, and Michele Papucci for the discussions on production via fragmentation in PYTHIA. The authors also thank Vsevolod Syvolap for reading the manuscript. The work of SM was produced by FermiForward Discovery Group, LLC under Contract No. 89243024CSC000002 with the U.S. Department of Energy, Office of Science, Office of High Energy Physics.

\newpage
\appendix
\onecolumngrid 

\section{Mixing angles}
\label{app:mixing}

In this section, we provide the expressions of the mixing angles between various LLPs and neutral mesons. 

\subsection{Vector LLPs}

The mixing of the dark photon $V$ with vector mesons $V^{0} = \rho^{0},\omega,\phi$ is similar to the mixings of the ordinary photon modulus of the tiny coupling $\epsilon$ in Eq.~\eqref{eq:lagrangian-vectors}. The mixing angles $\theta_{VV^{0}}$ can be derived using the approach of Hidden Local Symmetry of vector meson dominance~\cite{Fujiwara:1984mp} and have the form
\begin{equation}
    \theta^{\text{DP}}_{VV^{0}} = \frac{\sqrt{2}f_{\pi}e\epsilon \text{Tr}[T_{V_{0}}Q]}{m_{\rho}}\frac{m_{V_{0}}^{2}}{m_{V}^{2}-m_{V^{0}}^{2}-i\Gamma_{V^{0}}(m_{V})m_{V^{0}}},
    \label{eq:mixing-angles-dark-photon}
\end{equation}
with $e = \sqrt{4\pi\alpha_{\text{EM}}}$ being the Coulomb constant in natural units. Here, $T_{V^{0}}$ is the generator of the vector mesons,
\begin{equation}
    T_{\rho^{0}} = \frac{1}{2}\text{diag}(1, -1, 0), \quad T_{\omega} = \frac{1}{2}\text{diag}(1, 1, 0), \quad T_{\phi} = \frac{1}{\sqrt{2}}\text{diag}(0,0,1)
\end{equation}
while
\begin{equation}
    Q = \frac{1}{3}\text{diag}(2,-1,-1)
\end{equation}
is the generator of the electric charge. Finally, $\Gamma_{V^{0}}(s)$ is the scale-dependent decay width of the meson. In practice, this dependence is only relevant for broad resonances such as $\rho^{0}$, for which we use~\cite{Guo:2011ir}
\begin{equation}
    \Gamma_{\rho^{0}}(s) = \frac{\sqrt{s}}{m_{\rho^{0}}}\left(\frac{s-4m_{\pi}^{2}}{m_{\rho^{0}}^{2}-4m_{\pi}^{2}} \right)^{\frac{3}{2}}\Gamma_{\rho^{0}}(m_{\rho^{0}})
\end{equation}
The mixing angles for the $B-L$ mediators can be obtained using the mixing angles of dark photons using the relation~\cite{Tulin:2014tya,Ilten:2018crw}
\begin{equation}
    \theta^{\text{B-L}}_{VV^{0}} = \frac{\sqrt{4\pi\alpha_{B}}}{e\epsilon}\frac{\text{Tr}[T_{V^{0}}B]}{\text{Tr}[T_{V^{0}}Q]}\theta^{\text{DP}}_{VV^{0}},
\end{equation}
where $B = 1/3 \text{diag}(1,1,1)$ is the generator of the baryon current.

\subsection{ALPs}

We follow Ref.~\cite{Ovchynnikov:2025gpx} for deriving the interactions of ALPs with various mesons. Using the definitions of the ALP couplings $f,c_{u,d,s},c_{G}$ from the Lagrangian~\eqref{eq:lagrangian-alp}, introducing the parameter $\epsilon = f_{\pi}/f_{a}$, and calculating all the quantities in the limit $\mathcal{O}(\delta,\epsilon)$, where $\delta = (m_{d}-m_{u})/(m_{d}+m_{u})$ is the isospin breaking parameter, the ALP-meson mixing angles can be given in the form
\begin{align}
    \theta_{\pi^{0}a}&= \epsilon\bigg[\frac{\delta\frac{ m_{\pi^{0}}^2}{3}\left[\frac{m_a^2 \left(c_u+2 c_{s}\right)+m_a^2 c_d}{m_a^2-m_{\eta'}^2}-\frac{2\left(m_a^2 \left(c_{s}-c_{u}\right)-m_a^2 c_d\right)}{m_a^2-m_{\eta}^2}\right]-m_a^2
   \left(c_u-c_d\right)}{m_a^2-m_{\pi^{0}}^2}\\ &+ \delta c_{G}\frac{m_{\pi^{0}}^{2}}{3} \frac{\left(\frac{\left(2 m_a^2-m_{\eta'}^2-m_{\pi^{0}}^2\right)}{m_a^2-m_{\eta'}^2}-\frac{2 \left(m_a^2-2 m_{\eta}^2+m_{\pi^{0}}^2\right)}{m_a^2-m_{\eta}^2}\right)}{m_a^2-m_{\pi^{0}}^2}+ c_G \left(\kappa_d-\kappa_u\right)\bigg],
   \label{eq:mixing-pi0-1}
\\
\theta_{\eta a} &= \sqrt{\frac{2}{3}}\epsilon \left[-\frac{m_a^2 \left(c_d-c_s+c_u+\delta \frac{  m_{\pi^{0}}^2(c_d-c_u)}{m_a^2-m_{\pi^{0}}^2} \right)}{m_a^2-m_{\eta}^2}+\frac{c_G \left(m_a^2+m_{\pi^{0}}^2-2 m_{\eta
   }^2\right)}{m_a^2-m_{\eta}^2}-2 c_G \left(\kappa_d+\kappa_u\right)\right],
   \label{eq:mixing-eta-1}
\\
    \theta_{\eta' a} &=\sqrt{\frac{1}{3}}\epsilon  \left[-\frac{m_a^2 \left( c_d +2 c_s+c_u-\delta\frac{m_{\pi^{0}}^2 (c_d-c_u)}{m_a^2-m_{\pi^{0}}^2} \right)}{m_a^2-m_{\eta'}^2}-c_G \frac{\left(2 m_a^2-m_{\eta'}^2-m_{\pi^{0}}^2\right)}{m_a^2-m_{\eta'}^2}+c_G \left(\kappa_d+\kappa_u\right)\right]
    \label{eq:mixing-etapr-1}
\end{align}
Each of the mixing angles $\theta_{m^{0}a}$ includes the poles not only from the meson $m^{0}$ but also from the other mesons $m^{0'} = \pi^{0},\eta,\eta' \neq m^{0}$. It is because the minimal ChPT Lagrangian contains a mixing between $\pi^{0}$ and $\eta,\eta'$, which vanishes in the limit $\delta \to 0$. 

For completeness, we also provide the mixing angles between $\pi^{0}$ and $\eta/\eta'$ (modulus $\delta$) originating from the pure ChPT:
\begin{equation}
\theta_{\pi^{0}\eta} = -\frac{\sqrt{\frac{2}{3}} m_{\pi^{0}}^2}{m_{\eta}^2-m_{\pi^{0}}^2}, \quad  \theta_{\pi^{0}\eta'} = -\frac{m_{\pi^{0}}^2}{\sqrt{3} \left(m_{\eta'}^2-m_{\pi^{0}}^2\right)}
\label{eq:mixing-ChPT}
\end{equation}

\begin{table}[h!]
    \centering
    \begin{tabular}{|c|c|c|c|c|}
    \hline
 \text{Model} & $c_{u}$ & $c_{d}$ & $c_{s}$ & $c_{G}$ \\ \hline 
 BC10 & 0.873891 & 0.986109 & 0.986109 & 0.   \\  \hline
 BC11 & -0.0166131 & -0.0166131 & -0.0166131 & 1. \\ \hline
    \end{tabular}
    \caption{The renormalization group flow of the ALP couplings $c_{u},c_{d},c_{s},c_{G}$ from the scale $\Lambda = 1\text{ TeV}$, at which the ALP Lagrangian~\eqref{eq:lagrangian-alp} is defined, to the scale of interest $\Lambda_{Q} \simeq m_{a}$, for the models of ALPs universally coupled to fermions (BC10) and those coupled to gluons (BC11), see text for definition. The flow has been computed using the \texttt{Mathematica} notebook from Refs.~\cite{DallaValleGarcia:2023xhh,Ovchynnikov:2025gpx}.}
    \label{tab:rg-couplings}
\end{table}

The next step is to fix the couplings $\xi = \{c_{u},c_{d},c_{s},c_{G}\}$. The Lagrangian~\eqref{eq:lagrangian-alp} is defined above the electroweak scale $\Lambda > m_{t} = 173\text{ GeV}$, and hence at the low scales $\mu \simeq m_{a}$ of our interest, the effective coefficients $\xi$ experience non-trivial dynamics governed by the renormalization group equations~\cite{Bauer:2020jbp}. In this study, we will consider the two ALP models: the one with universal coupling to fermions (the so-called BC10 model from~\cite{Beacham:2019nyx}), for which 
\begin{equation}
c_{q}(\Lambda = 1\text{ TeV}) = 1, \quad c_{G}(\Lambda = 1\text{ TeV}) = 0,
\end{equation}
and the one with the coupling to gluons (the BC11 model), with 
\begin{equation}
c_{q}(\Lambda = 1\text{ TeV}) = 0, \quad c_{G}(\Lambda = 1\text{ TeV}) = 1
\end{equation}
The list of the couplings and their values at $\mu \simeq 1\text{ GeV}$ is given in Table~\ref{tab:rg-couplings}.

\section{Production of the ALPs in the fragmentation}
\label{app:alp-fragmentation}

In this section, we discuss the effective mixing rate $\Theta_{m^{0}a}$ of ALPs produced in the quark fragmentation. We follow Ref.~\cite{Ovchynnikov:2025gpx}.

First, utilizing the effective interaction Lagrangian of ALPs with the lightest pseudoscalar mesons, we consider the following processes:
\begin{equation}
    \pi^{0}\pi^{0}\to \pi^{0}a, \quad \pi^{0}\eta \to \pi^{0}a, \quad \pi^{0}\eta' \to \pi^{0}a
\end{equation}
Their matrix elements have the form 
\begin{equation}
\mathcal{M}_{\pi^{0}m^{0}\to a \pi^{0}} = \Theta_{m^{0}a}\cdot \tilde{\mathcal{M}},
\end{equation}
where $\Theta_{m^{0}a} = \theta_{m^{0}a}+ \dots$. They are also the simplest scattering processes. Having calculated $\Theta_{m^{0}a}$, we extrapolate this coupling onto the generic fragmentation chain when the meson $m^{0}$ is replaced with the ALP.

To obtain the expression for $\Theta_{m^{0}a}$, we utilize the Mathematica notebook from Ref.~\cite{Ovchynnikov:2025gpx}. Explicitly, they read
\begin{multline}
\Theta_{\pi^{0}a}= \theta_{\pi^{0}a}+c_G \left[\kappa_u-\kappa_d-\delta  \left(\kappa_d+\kappa_u\right) \left(\sqrt{3} \theta_{\pi^{0}\eta'}+\sqrt{6} \theta_{\pi^{0}\eta}+1\right)\right]\\ -\frac{1}{3} \delta  \left(3 \theta_{\pi^{0}\eta'}+3 \sqrt{2}\theta_{\pi^{0}\eta}+\sqrt{3}\right) \theta_{\eta'a}-\frac{1}{3} \delta \theta_{\eta a} \left(3 \sqrt{2} \theta_{\pi^{0}\eta'}+6 \theta_{\pi^{0}\eta}+\sqrt{6}\right),
\label{eq:theta-eff-pi0a}
\end{multline}
\begin{multline}
\Theta_{\eta a}= \theta_{\eta a} +\frac{1}{2} c_G \left[\kappa_d \left(\delta  \left(2 \sqrt{2} \theta_{\pi\eta'}+\theta_{\pi\eta
   }+\sqrt{6}\right)+\sqrt{6}\right)+\kappa_u \left(\sqrt{6}-\delta  \left(2 \sqrt{2} \theta_{\pi\eta'}+\theta_{\pi\eta
   }+\sqrt{6}\right)\right)\right]\\ -\frac{1}{2} \delta  \theta_{\pi^{0}a} \left(2\sqrt{2}\theta_{\pi\eta'}+\theta_{\pi\eta}+\sqrt{6}\right) +\frac{\theta_{\eta'a}}{\sqrt{2}}
\label{eq:theta-eff-etaa}
\end{multline}
\begin{multline}
    \Theta_{\eta'a} = \theta_{\eta'a} + c_G \left[\kappa_d \left(\delta  \left(-\theta_{\pi^{0}\eta'}+2 \sqrt{2} \theta_{\pi^{0}\eta}+\sqrt{3}\right)+\sqrt{3}\right)+\kappa_u\left(\delta  \left(\theta_{\pi^{0}\eta'}-2 \sqrt{2} \theta_{\pi^{0}\eta}-\sqrt{3}\right)+\sqrt{3}\right)\right] +\\+ \delta \cdot\theta_{\pi^{0}a} \left(\theta_{\pi^{0}\eta'}-2 \sqrt{2} \theta_{\pi^{0}\eta}-\sqrt{3}\right)+\sqrt{2} \theta_{\eta a}
\label{eq:theta-eff-etapra}
\end{multline}
Here, Inserting the mixing angles~\eqref{eq:mixing-pi0-1}-\eqref{eq:mixing-etapr-1} and~\eqref{eq:mixing-ChPT} in the effective mixing angles~\eqref{eq:theta-eff-pi0a}-\eqref{eq:theta-eff-etapra}, we can explicitly verify that the unphysical $\kappa_{q}$ dependence drops out.

\section{Various contributions to neutral pion production}
\label{app:pion-production-from-decays}

\begin{table}[h]
    \centering
    \begin{tabular}{|l|c|c|c|}
        \hline
        Mother ID & SPSBD & LHC & FermilabBD \\
        \hline
          15 (Unknown) & 2 & 1259 & 0 \\
          113 ($\pi^0$) & 439 & 3710 & 299 \\
          213 ($\rho^0$) & 1018753 & 8715842 & 671834 \\
          221 ($\eta$) & 567420 & 4454086 & 387640 \\
          223 ($\omega$) & 535922 & 4378409 & 358151 \\
          225 (Unknown) & 2 & 72 & 0 \\
          310 ($K^0_S$) & 192561 & 1917421 & 116721 \\
          313 ($K^*(892)^0$) & 63436 & 692092 & 36402 \\
          323 ($K^*(892)^+$) & 70804 & 699926 & 43175 \\
          331 ($\eta'$) & 22879 & 205560 & 15046 \\
          333 ($\phi$) & 1560 & 17615 & 881 \\
          411 ($D^0$) & 32 & 19953 & 0 \\
          413 ($D^*(2010)^+$) & 15 & 15874 & 2 \\
          421 ($D^+$) & 108 & 88191 & 9 \\
          423 ($D^*(2010)^0$) & 27 & 32076 & 6 \\
          431 ($D_s^+$) & 2 & 2083 & 0 \\
          433 ($D_s^*(2112)^+$) & 1 & 781 & 1 \\
          443 ($J/\psi$) & 10 & 803 & 1 \\
          445 ($\psi(2S)$) & 43 & 1040 & 8 \\
          2114 ($\Delta^0$) & 93068 & 395243 & 80788 \\
          2214 ($\Delta^{++}$) & 130236 & 427349 & 121185 \\
          3114 ($\Sigma^{*0}$) & 699 & 5943 & 409 \\
          3122 ($\Lambda$) & 74230 & 498875 & 56066 \\
          3214 ($\Sigma^+$) & 18174 & 94904 & 14506 \\
          3222 ($\Sigma^0$) & 25572 & 208135 & 17799 \\
          3224 ($\Sigma^-$) & 1593 & 6953 & 1296 \\
          3314 ($\Xi^0$) & 455 & 5888 & 229 \\
          3322 ($\Xi^-$) & 8708 & 102372 & 4749 \\
          3324 ($\Xi^*$) & 634 & 6101 & 417 \\
          3334 ($\Omega^-$) & 19 & 272 & 14 \\
          4122 ($\Lambda_c^+$) & 9 & 3749 & 1 \\
          4212 ($\Sigma_c^0$) & 5 & 520 & 0 \\
          10441 ($h_c$) & 51 & 995 & 11 \\
          20213 ($a_0^0$) & 7 & 6391 & 0 \\
          20313 ($a_2^0$) & 5 & 1138 & 1 \\
          20443 ($\chi_{c1}$) & 0 & 335 & 3 \\
          Fragmentation & 1295015 & 9239445 & 919676 \\
        \hline
        Total & 4122498 & 32260346 & 2847326 \\
        \hline
    \end{tabular}
    \caption{Various neutral pion production sources from different facilities.}
    \label{tab:pions}
\end{table}

Table~\ref{tab:pions} shows the contributions to the fluxes of the lightest pseudoscalar meson $m^{0} = \pi^{0}$ as obtained by running \texttt{PYTHIA8} with the setups corresponding to the SPSBD, FermilabBD, and LHC facilities. 

The results indicate that practically independently of the facility, around 60-70\% of the pions originate from decays of heavier mesons, whereas the rest originate from the fragmentation process.

\section{\texttt{PYTHIA8} repository}
\label{app:pythia-run}

In this appendix, we present our modification of the \texttt{PYTHIA8} code, available at \url{https://gitlab.com/YehorKyselyov/pythia-mixing/-/tree/dev}. Our main modification contribution lies in introducing fragmentation and meson decay channels, along with example programs to accurately simulate production in proton-proton collisions.

\subsection{Fragmentation in \texttt{PYTHIA8}}
To implement production by fragmentation in \texttt{PYTHIA8}, we modified the source code responsible for the fragmentation chain, specifically \texttt{FragmentationFlavZpT.cc}. In the \texttt{StringFlav::combine} method, which combines two flavors (including diquarks) to produce a hadron, if the resulting particle ID corresponds to a neutral meson, the \texttt{StringFlav::MixHiddenValleyMesons} method is called. This method checks several flags, defined in a set of meson mixing configurations. For each meson type, the method verifies if the corresponding configuration flag (e.g., \texttt{HVparams:mixWithPi0}, \texttt{HVparams:mixWithEta}, etc.) is enabled. If the flag is set and the mixing rate, defined by the associated rate parameter (e.g., \texttt{HVparams:mixingRateWithPi0}, \texttt{HVparams:mixingRateWithEta}, etc.), is met based on a random number between 0 and 1, the meson ID is replaced with that of a hidden valley meson, which represents the given LLP in our case.

Let us demonstrate how to enable meson mixing for the dark photon model. Specifically, we enable mixing with the $\rho_0$, $\omega$, and $\phi$ mesons. The following code shows how to apply this mixing to \texttt{PYTHIA8}:

\begin{lstlisting}[language=C++]
void ApplyDPFragmentation(PYTHIAParallel &pythia, double mass)
{
    // Enable Mixing for Specific Mesons
    pythia.readString("HVparams:mixWithRho0 = true");
    pythia.readString("HVparams:mixWithOmega = true");
    pythia.readString("HVparams:mixWithPhi = true");

    // Set Mixing Rates
    pythia.readString("HVparams:mixingRateWithRho0 = " + std::to_string(MixingRho0(mass)));
    pythia.readString("HVparams:mixingRateWithOmega = " + std::to_string(MixingOmega(mass)));
    pythia.readString("HVparams:mixingRateWithPhi = " + std::to_string(MixingPhi(mass)));

    // Set Hidden Valley Meson Mass
    pythia.readString("4900111:m0 = " + std::to_string(mass));
    pythia.readString("4900221:m0 = " + std::to_string(mass));
    pythia.readString("4900331:m0 = " + std::to_string(mass));
}
\end{lstlisting}

When the simulation runs, mesons with IDs 111 (neutral pion), 221 (eta), and 331 (eta prime) will be replaced with hidden valley mesons (IDs 4900111, 4900221, and 4900331) at rates determined by the computed mixing functions (\texttt{MixingRho0(mass)}, \texttt{MixingOmega(mass)}, \texttt{MixingPhi(mass)}), as defined in Eq.~\eqref{eq:theta-eff}. This approach allows the simulation of dark photon mixing with these mesons, enabling the study of their impact on particle production and interactions in \texttt{PYTHIA8}.

\textbf{NOTE}: Flags \texttt{mixWithRho0} and others are default to \texttt{false}, meaning that if they are not specified, PYTHIA will behave as usual.   
\subsection{Decays in \texttt{PYTHIA8}}

To implement decay channels for neutral mesons in \texttt{PYTHIA8}, we use the built-in \texttt{addChannel} method. This method allows adding a decay mode for a given particle ID with the following syntax:

\begin{lstlisting}
    id:addChannel = onMode bRatio meMode product1 product2 ...
\end{lstlisting}

Let us demonstrate how to enable production from neutral mesons for the dark photon model:
\begin{lstlisting}[language=C++]
void ApplyDPDecays(PYTHIAParallel &pythia, double mass)
{
    pythia.readString("111:addChannel = 1 " + std::to_string(BranchingPi0(mass)) + " 0 22 4900111");
    pythia.readString("221:addChannel = 1 " + std::to_string(BranchingEta(mass)) + " 0 22 4900221");
    pythia.readString("331:addChannel = 1 " + std::to_string(BranchingEtaPrime(mass)) + " 0 22 4900331");
}
\end{lstlisting}
Here, the functions \texttt{BranchingPi0(mass)}, \texttt{BranchingEta(mass)}, and \texttt{BranchingEtaPrime(mass)} compute the branching ratios for the decays, as defined in Eq.~\eqref{eq:br-DP}. Each command enables a new decay channel where mesons (\(\pi^0\), \(\eta\), and \(\eta'\)) decay into a photon and a hidden valley meson (\(4900111\), \(4900221\), \(4900331\)).

% Overall, to calculate B-L branching ratios, we use Eq.~\eqref{eq:br-BL}. For axion-like particles (ALPs), we followed the approach described in Sec.~\ref{sec:approach} and implemented the branching fractions in Wolfram Mathematica. Since the analytical expressions were quite large, we tabulated the results and used interpolation in our C++ programs to compute the branching ratios.
\subsection{Description of \texttt{PYTHIA8} Example Programs}

In the \texttt{example} folder, several new files are present, namely:
\begin{itemize}[label=-]
    \item \texttt{MixingPattern.cc} and \texttt{MixingPattern.h}
    \item \texttt{FacilitySettings.cc} and \texttt{FacilitySettings.h}
    \item \texttt{Interpolation.cc}, \texttt{Interpolation.h} and several \texttt{.csv} files containing tabulated functions for branching ratios and mixing angles as a function of mass for ALPs models.
    \item \texttt{mixing01.cc} program simulates events and provides output describing X particle kinematics.
    \item \texttt{mixing02.cc} program simulates the yield of new X particles, as described by \eqref{eq:production-probability}. While the same result can be obtained by analyzing the output of \texttt{mixing01.cc}, this program provides a quicker and more efficient approach.
\end{itemize}

Below, we discuss them in detail.

\subsubsection{Overview of \texttt{MixingPattern.h}}
An abstract class, \texttt{MixingPattern}, serves as the base for different mixing models. This class is extensively used throughout our programs.

The class serves two main purposes: first, to apply the corresponding mixing to a \texttt{PYTHIAParallel} object; and second, to calculate the corresponding \textbf{weight} for an event. Since the mixing rate can be large, we use a "fake" mixing rate in PYTHIA simulations. After an event occurs, the \texttt{CalculateWeight} method adjusts for this by returning the correct mixing angle for fragmentation production or the correct branching ratio for neutral meson decay channels. These calculations are based on the equations derived in Sec.~\ref{sec:approach}.

For vector models, the expressions are used directly. For ALPs, we implemented the branching fractions in Wolfram Mathematica. Due to the complexity of the analytical expressions, we tabulated the results and used interpolation in our C++ programs to efficiently compute the branching ratios.

\begin{lstlisting}[language=C++]
class MixingPattern {
public:
    MixingPattern(const std::string &name, double mixingRate);
    virtual ~MixingPattern() = default;

    virtual void ApplyMixing(PYTHIA8::PYTHIAParallel &pythia, double mass) const = 0;
    virtual double CalculateWeight(double mass, int hiddenValleyId) const;
    virtual double CalculateWeight(double mass, const PYTHIA8::Event &event, int hvIndex) const;
    virtual std::string GetProductionType(int motherId) const = 0;

    std::string GetName() const;

protected:
    double mixingRate_;
    std::string name_;
};
\end{lstlisting}

We then implement derived classes that handle the specifics of mixing, including decays and meson fragmentation processes. The models included in the latest version of the code correspond to those discussed in this paper, namely:  
\begin{itemize}[label=-]
    \item Dark Photon model (\texttt{class DPMixing})  
    \item B-L mediators (\texttt{class BLMixing})  
    \item ALP coupled to gluons (\texttt{class BC11})  
    \item ALP universally coupled to fermions (\texttt{class BC10})  
\end{itemize}

\subsubsection{Overview of \texttt{FacilitySettings.h}}
Another key abstraction in our programs is the \texttt{FacilitySettings} class, which is responsible for applying the experimental settings corresponding to the experiment of interest. This class configures PYTHIA based on the specific requirements of each experimental setup, such as beam parameters, detector settings, and other facility-specific configurations.

The primary purpose of this class is to encapsulate and apply experiment-specific configurations (tunes) to a \texttt{PYTHIAParallel} object. By centralizing these settings, \texttt{FacilitySettings} ensures consistency and modularity in the simulation environment, making it easy to adapt to different experimental conditions.
\begin{lstlisting}[language=C++]
class FacilitySettings {
public:
    FacilitySettings(const std::string &name, int beamEnergyA, int beamEnergyB, bool softQCDOn, int pomFlux, int pdfSet);
    virtual ~FacilitySettings() = default;

    virtual void applySettings(PYTHIA8::PYTHIAParallel &pythia) const;
    std::string GetName() const;

protected:
    std::string name_;
    int beamEnergyA_;
    int beamEnergyB_;
    bool softQCDOn_;
    int pomFlux_;
    int pdfSet_;
};
\end{lstlisting}

We then implement derived classes that handle specific configurations for various experiments. Cases covered in the present version of the code correspond to those discussed in this paper, namely:
\begin{itemize}[label=-]
    \item CERN LHC for different choices of the forward LHC tune from Ref.~\cite{Fieg:2023kld}. The following settings represent variations in the beam remnant parameters:
    \begin{itemize}
        \item \texttt{LHCSettingsMiddle}: A baseline tune with \(\texttt{primordialKTsoft} = 0.58\) GeV and \(\texttt{primordialKTremnant} = 0.58\) GeV.
        \item \texttt{LHCSettingsHigh}: A higher tune variation where \(\texttt{primordialKTsoft}\) and \(\texttt{primordialKTremnant}\) are increased to 1.27 GeV.
        \item \texttt{LHCSettingsLow}: A lower tune variation with \(\texttt{primordialKTsoft} = 0.26\) GeV and \(\texttt{primordialKTremnant} = 0.26\) GeV.
        \item \texttt{LHCSettingsMixed1}: A mix where \(\texttt{primordialKTsoft} = 1.27\) GeV and \(\texttt{primordialKTremnant} = 0.26\) GeV.
        \item \texttt{LHCSettingsMixed2}: A mix where \(\texttt{primordialKTsoft} = 0.26\) GeV and \(\texttt{primordialKTremnant} = 1.27\) GeV.
    \end{itemize}  
    \item Fermilab (\texttt{FermilabBDSettings})  
    \item CERN SPS (\texttt{SPSBDSettings})  
    \item Serpukhov Accelerator (\texttt{SerpukhovBDSettings})  
    \item Future Circular Collider (\texttt{FCChhSettings})  
\end{itemize}

% {\color{red}Specify what are these \texttt{LHCSettingsMiddle} etc.}

\subsubsection{\texttt{Interpolation.h}}  

The \texttt{Interpolation.h} file handles linear interpolation, to compute the correct weights for ALPs models. It provides methods for interpolating data from predefined tabulations:  

\begin{lstlisting}
    double interpolate(const std::vector<double> &xPoints, const std::vector<double> &yPoints, double x);
    double InterpolateFromFile(const std::string &filename, int valueColumn, double ma);
\end{lstlisting}  

The relevant tabulated data -- the branching ratios of the ALPs production in decays of mesons, the squared effective mixing angles $|\Theta_{m^{0}a}|^{2}$ (all assuming the coupling $f_{\pi}/f_{a} = 1$) -- is stored in CSV files, including:  

\begin{itemize}[label=-]
    \item \texttt{BrRatios-Fermion-universal-Lambda-1.-TeV.csv}
    \item \texttt{theta2\_eff-Fermion-universal-Lambda=1.-TeV.csv}
    \item \texttt{BrRatios-Gluon-Lambda-1.-TeV.csv}
    \item \texttt{theta2\_eff-Gluon-Lambda=1.-TeV.csv}
\end{itemize}

Similar files for the ALPs other than implemented in our code by default may be produced by the \texttt{Mathematica} notebook associated with the paper~\cite{Ovchynnikov:2025gpx}. This way, incorporating various ALPs is very simple: having the tabulated data, one just needs to create a dedicated new ALP class similar to the already present classes \texttt{BC10} and \texttt{BC11}.

\subsubsection{Overview of \texttt{mixing01.cc}}

Now, we take a closer look at The \texttt{mixing01.cc} program is designed to provide a detailed description of the kinematics and production of LLPs. The program accepts five command-line arguments:
\begin{enumerate}
    \item The mass of the new physics particle,
    \item The number of proton-proton collisions to simulate,
    \item The name of the facility,
    \item The name of the tune,
    \item The new physics model name,
    \item The output directory for the results.
\end{enumerate}

The tune corresponds to the \texttt{PYTHIA8} setup that is used to simulate the 

Based on the input, the program fetches the corresponding \texttt{FacilitySettings} and \texttt{MixingPattern} objects using the \texttt{GetFacilitySettings} and \texttt{GetMixingPattern} functions. The \texttt{SetUpPYTHIAInstance} function is called to configure the PYTHIA simulation with these settings.

The \texttt{SimulateEventsAndCountParticles} function simulates the specified number of events. Within each event, every implemented production mode (fragmentation, decays of mesons) is assumed to produce the LLP with the seed probability $P_{\text{seed}}=0.01$. Then, the program iterates over all final states and checks for the presence of the particles with the ID starting with "4900", belonging to the Hidden Valley Particle but corresponding to different LLPs in our model.

The output is a text file in scientific notation with six decimal places, which is saved in the specified output directory. It contains the following columns: production type, ID of the hidden valley particle, production weights $P_{\text{true}}$, mass, energy, and the polar angle of the particle. 

For the fragmentation production type, the ID ending with \texttt{XXX} (111, 221, 331, 113, 223, or 333) specifies the mixing of the meson which is responsible for the production. 

The weight is equal to
\begin{equation}
    P_{\text{true}}=\frac{1}{N_{\text{simulated}}\cdot P_{\text{seed}}} \times \begin{cases}
    |\Theta_{m^{0}X}|^{2}, \quad \text{fragmentation}, \\ \text{Br}(m\to X), \quad \text{decays},
    \end{cases}
\end{equation}
where the effective squared mixing rate $|\Theta_{m^{0}X}|^{2}$ and the branching ratio $\text{Br}(m\to X)$ are evaluated at the unit value of the LLP coupling, defined in the Lagrangians~\eqref{eq:lagrangian-vectors},~\eqref{eq:lagrangian-alp} ($\epsilon$ for dark photons, $\alpha_{B}$ for the mediators coupled to the baryon current, and $f_{\pi}/f_{a}$ for the ALPs).

The total number of events with LLPs per proton collision can be obtained by summing all the weights, while the angle-energy distribution can be obtained by weighting the values of the angles and energies of the LLPs.

\subsubsection{Overview of \texttt{mixing02.cc}}

The \texttt{mixing02.cc} program is designed to calculate the probability of LLP production in a single proton-proton collision. It takes four command-line arguments:
\begin{enumerate}
    \item The mass of the LLP,
    \item The name of the facility,
    \item The name of the tune,
    \item The model,
    \item The output directory for the results.
\end{enumerate}

Based on the input, the program fetches the corresponding \texttt{FacilitySettings} and \texttt{MixingPattern} objects using the \texttt{GetFacilitySettings} and \texttt{GetMixingPattern} functions. The \texttt{SetUpPYTHIAInstance} function is called to configure the PYTHIA simulation with these settings.

Similarly to the \texttt{mixing01.cc} program, \texttt{mixing02} uses these inputs to simulate the production of LLPs in \texttt{pp} collisions and calculate the production probability for the specified LLP mass and model. 

The difference is that the output only includes the LLP overall production probabilities, which are written to small output files. It may be used when one is interested solely in the integrated LLP flux, being agnostic about the angle-energy distribution. The output files are named according to the tune, model, and mass of the LLP for easy identification and further analysis.

\newpage 

\bibliography{bib.bib}

\end{document}